\documentclass[%
 reprint,
 amsmath,amssymb,
 aps,
]{revtex4-2}

\usepackage{hyperref}
\usepackage{graphicx}
\usepackage{dcolumn}
\usepackage{bm}
\usepackage{subfig}
\usepackage{color,soul}

\begin{document}

\preprint{APS/123-QED}

\title{Resonant excitation of very high gradient plasma wakefield accelerators by optical-period bunch trains}

\author{P. Manwani}
\email{pkmanwani@gmail.com}
\author{N. Majernik}
 
\author{J. B. Rosenzweig}%

\affiliation{ 
Department of Physics and Astronomy, UCLA, Los Angeles, California 90095}%

\date{\today}

\begin{abstract}

Using a periodic electron beam bunch train to resonantly excite plasma wakefields in the quasi-nonlinear (QNL) regime has distinct advantages over employing a single, higher charge bunch. Resonant excitation in the QNL regime can produce plasma electron blowout using very low emittance beams with a small charge per pulse: the local density perturbation is extremely nonlinear, achieving total rarefaction, yet  the resonant response of the plasma electrons at the plasma frequency is preserved. Such a pulse train, with inter-bunch spacing equal to the plasma period, can be produced via inverse free-electron laser bunching. To achieve resonance with a laser wavelength of a few microns, a high plasma density is used, with the attendant possibility of obtaining extremely large wakefield amplitude, near 1 TV/m for FACET-II parameters. In this article, we use particle-in-cell simulations to study the plasma response, the beam modulation evolution, and the instabilities encountered, that arise when using a bunching scheme to resonantly excite waves in a dense plasma.

\end{abstract}

\maketitle

\section{\label{sec:level1}INTRODUCTION}

In a beam driven plasma wakefield accelerator, electromagnetic fields are excited by an intense, relativistic particle beam driver. An oscillating plasma wave trails behind the driver and these fields can be utilized to accelerate electrons. The trailing, accelerating electron bunch is termed a witness beam. The use of a pulse train instead of a single bunch as the driver to resonantly excite the plasma wakefield permits new methods of efficiency enhancement. For example, in this scenario, the driving and accelerating beams may be interleaved, giving the opportunity to strongly enhance the beam loading and the energy extraction from the wave. 

The pulse trains needed for this resonant drive scheme can be produced through inverse free-electron laser (IFEL) bunching, as demonstrated recently through experiments performed at the Linac Coherent Light Source (LCLS) at the SLAC National Accelerator Laboratory \cite{exleap}. Resonant excitation requires stable, frequency-locked wakes which, given the presence of wave-breaking and amplitude-dependent wave frequency, is not straightforward in the strongly nonlinear ``blowout" regime \cite{quasi,sam_thesis}. Nevertheless, resonant excitation by trains with bunch-to-bunch spacing on the order of femtoseconds in the particularly advantageous {\sl quasi-nonlinear regime} (QNL) is investigated here. In this regime, plasma electron blowout is achieved, but with very low emittance beams self-consistently possessing  small charge per pulse. In the QNL regime, the plasma electrons are rarefied from the beam path, as in the blowout regime, a quite nonlinear process arising from the strong local charge density perturbation employed, that is the plasma is underdense - with beam density much larger than that of the plasma, $n_b\gg n_0$. In QNL operation, the beam is confined to a radial region significantly smaller than the plasma wavelength, $2 \pi k_p^{-1}$, where $k_p=\omega_p/c=\sqrt{4\pi r_e n_0}$. In this case, the global plasma wave disturbance is governed by a nearly \textit{linear} frequency response. The plasma waves may thus still be excited in a resonant fashion through bunch trains that have the same period as the linear plasma frequency.


Techniques for creating electron bunch trains are an active area of investigation. Phase space masking techniques exploiting transverse dispersion of energy-chirped beams can produce trains with inter-bunch spacing on the order of 100's of $\mu$m, as demonstrated by recent QNL regime experiments at the BNL ATF \cite{experiments,sam_thesis}.  However, to produce bunch spacing of a few $\mu$m, an IFEL-based approach is much more feasible. These tightly spaced trains are motivated by the desire to obtain very high fields. For example, a train with bunch-to-bunch spacing of 2 $\mu$m is resonant with a very high density plasma:  $n_0$ = ${k_{p}^{2}}/{4 \pi r_e}$ = 2.79 $\times$ $10^{20}$ cm$^{-3}$. This high density, in combination with assumptions of quasi-nonlinear waves approaching wave-breaking amplitude, $E_\mathrm{WB}=m_ec^2k_p$, implies extremely large-amplitude excited wakefields, up to TV/m, as discussed below.

Experimentally, a cryogenically-cooled gas jet operated at many-atmospheres of pressure may be used to produce the required density \cite{denseplasma}. Matching the beam into such a dense plasma \cite{match} requires an extremely short focusing beta-function: $\beta_{eq}$=$\sqrt{2 \gamma} k_p^{-1}$ $< 100$ $\mu$m for the 10 GeV beams foreseen at FACET-II \cite{FACETII}. This is a key challenge in the experimental realization of this scenario. A very high gradient (up to 700 T/m) permanent magnet quadrupole triplet \cite{magnet} can begin to focus the beam into the plasma, with further focusing by an adiabatic ramping \cite{adiabatic} of plasma density necessary to achieve the final, matched beta-function. For example, a 3.2 pC beam with a normalized emittance of 50 nm-rad \cite{topgunRosenzweig} has a matched spot size, $\sigma_x$, of 13 nm; comparable to a linear collider final focus \cite{finalfocus}. At this size, the beam creates enormous radial electric fields, exceeding 1 TV/m, which will ionize the gas atoms in a high field process termed the barrier suppression regime \cite{coulomb}. The beam parameters described are expected to be produced at SLAC's FACET-II facility \cite{FACETII} although modifications to the photoinjector employed will be necessary \cite{topgunCahill}, as well as an upgrade to the final focus system with high gradient quadrupole magnets, and the implementation of an IFEL buncher \cite{xleap}. This set of experiments has been formally approved for FACET-II with the experimental number E-317 assigned. This article employs QuickPIC simulations \cite{huang} to explore both the general characteristics of this scheme as well as the specific case of E-317 experimental parameters.

\section{PERIODIC BUNCHING WITH IFEL}

First proposed in 2004 \cite{sase}, the inverse free electron laser (IFEL) technique for micro-bunching (periodic bunching at the optical scale), was suggested as a means of generating high peak current bunch trains. As the initial foreseen application was in SASE FELs, this process has also been termed ``enhanced self-amplified spontaneous emission'' (ESASE) bunching. Here, the same approach may be used to produce the optical-period bunch trains required for driving resonant beam-plasma interactions, while maintaining excellent control over the beam quality \cite{duris}. In this technique, the electron peak current is significantly increased by an interaction between the electron beam and a high peak power laser pulse in a magnetic wiggler. The wiggler period, $\lambda_w$, and wiggler parameter, $K_w$ = $eB_{w} \lambda_{w}/(2 \pi m_e c)$, where $B_w$ is the peak magnetic field, are chosen such that $\lambda_{L}$ = $\lambda_{w} (1 + K_{w}^{2}/2)/2 \gamma_{b}^{2}$, where $\lambda_L$ is the laser wavelength and $\gamma_b$ is the relativistic factor for the average beam energy. An energy modulation at the laser wavelength $\lambda_L$ is imparted on the electron beam, which is then converted into a density modulation in a magnetic chicane \cite{xleap}.

Using this method to produce the pulse train desired for a range of applications has numerous advantages. Since the process can take place at relatively higher energy, space-charge induced emittance growth is suppressed. Further, since it takes place in a vacuum there is less degradation due to wakefield interactions; finally, due to the relatively modest bending needed, the effects of coherent synchrotron radiation (CSR) are mitigated \cite{microbunch}. This has been demonstrated  with a 2 $\mu$m laser in  the X-ray Laser Enhanced Attosecond Pulse Generation (XLEAP) project \cite{exleap}, illustrating the feasibility of micro-bunch creation in a scenario very close to that which is foreseen for FACET-II experiments. 

\section{\label{sec:level2}STUDY OF RESONANT PLASMA WAKEFIELD EXCITATION}

In a plasma wakefield accelerator, the plasma is initially set into motion by the forces associated with the electromagnetic fields of the driving particle bunch. In the linear regime, the resultant perturbation of the plasma density is small compared to $n_0$. In contrast, if the beam is very dense, the plasma electrons are completely ejected from the beam channel, leaving a ``bubble" in the density profile. In this case, one refers to the nonlinear (blowout) regime. Unlike in the linear regime, with its non-ideal, radially and temporally varying focusing fields, the blowout regime possesses a radial focusing field that is linear in $r$ and constant along the length of the bubble \cite{quasi}; the focusing is emittance-preserving. Further, since the magnitude of its accelerating field is independent of radial position \cite{blowout} in the blowout regime, there is no introduction of radially-dependent energy spread, in contrast to the linear regime. These aspects of the transverse and longitudinal wakefields, respectively, facilitate the acceleration of high quality beams over long distances. 

In general, however, the period of the excited plasma wave in the blowout regime is dependent on the charge density of the bunches, unlike the linear regime which depends only on the plasma density and beam axial velocity. Indeed, in the linear regime, the plasma response can be resonantly excited by a pulse train, with each bunch adding to the plasma wave via constructive superposition. Resonant excitation in the linear regime has been experimentally demonstrated by modulating an electron beam into a train of micro-bunches spaced at a laser wavelength of 10.6 $\mu$m through an IFEL interaction at the BNL ATF facility \cite{resonant}. This resonant excitation process may continue until the non-linear regime is approached, and the resonance breaks down due to the amplitude-dependent expansion of the plasma wave period \cite{blowout}. The quasi-nonlinear regime, on the other hand, exploits the advantages of both the linear and non-linear regimes by employing low emittance, tightly focused beams with relatively small charges. In this scenario, the local beam density can greatly exceed that of the surrounding plasma, while simultaneously having a smaller total charge than the relevant plasma electrons taking part in the interaction, thus allowing for blowout while still maintaining a quasi-linear frequency response in the bulk plasma \cite{Barber}.  We expand on this explanation in what follows.

\begin{figure}
    \centering
    \includegraphics[width=2.8in,height=4in]{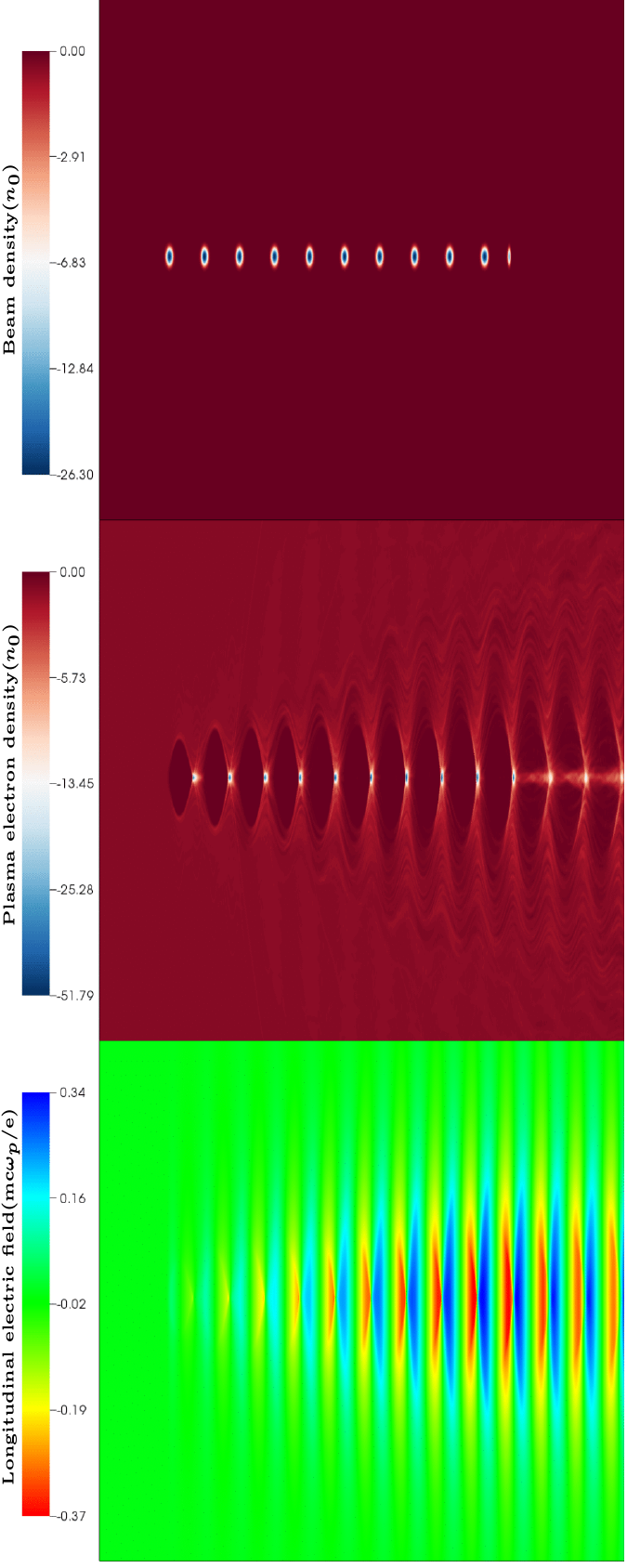}
    \caption{PIC simulation snapshots of the beam charge density, plasma electron density, and longitudinal electric field for the ten bunches followed by a witness bunch (charge per driver bunch = 0.32 pC).}
    \label{fig:10bunch}
\end{figure}

The relevant plasma charge taking part in the interaction is quantified by the number of electrons contained in a cubic plasma skin depth, $n_0 k_p^{-3}$. The parametric limits for this resonant excitation in the QNL regime are thus characterized by the normalized charge density $\widetilde{Q}$ \cite{physics/0205007}:

\begin{equation}
    \widetilde{Q} = m \frac{N_b k^{3}_p}{n_0} =  m 4 \pi r_{e} k_{p} N_b
    \label{eq:q}
\end{equation}

\noindent where the factor $m$ is the number of bunches in the train, $N_b$ is the number of electrons in each bunch, and $r_e$ is the classical electron radius. This parameter is the beam charge normalized to the relevant plasma electron charge. Indeed, in order to utilize this definition, each microbunch must be assumed to occupy a volume smaller than approximately $k_p^{-3}$, simultaneously obeying the conditions $k_p\sigma_x \ll 1$ and $k_p\sigma_z < 1$; this second criterion is already implied by the assumption of a resonant pulse train yielding efficient excitation. 

The condition for the wakefield to be definitively in the blowout regime, in the case of a single bunch, is met when $\widetilde{Q}>1$ which implies that $n_b>n_0$, independently of the values of $k_p\sigma_z$ and $k_p\sigma_x$, as long as they are less than unity. The parameter $\widetilde{Q}$ can, on the other hand, with knowledge of these beam parameters, be taken as a measure of the nonlinearities present in the beam-plasma interaction \cite{physics/0205007}, such as period lengthening and wave steepening. With a resonant bunch train, the wake responses of the bunches should add linearly, so the total charge of the train is used to calculate $\widetilde{Q}$, inspiring the introduction of the factor $m$ above. This superposition is expected to hold true in the case of linear, resonant excitation, and also in the QNL regime. In the present study, the limits on resonance given $\widetilde{Q}$ are explored. In particular, nonlinear detuning of the resonant frequency is expected as $\widetilde{Q}$ approaches unity, and the QNL regime is exceeded in favor of the blowout regime.

The PIC simulations employed in this paper are carried out using a plasma density, $n_0=2.79 \times 10^{20}$ cm$^{-3}$, corresponding to a plasma wavelength, $\lambda_p$, of 2 $\mu$m. Significantly, this density also implies a wave-breaking field (a guideline to the scaled of the maximum wakefield amplitude obtainable), of $E_\mathrm{WB}=1.6$ TV/m. The electron beams are assumed to have transversely symmetric Gaussian profiles with an energy of 10 GeV, consistent with FACET-II expectations. The charge and emittance for the micro-bunches are consistent with the results of Ref.  \cite{topgunRosenzweig}, with each micro-bunch having a charge of 0.32 pC and 50 nm-rad normalized emittance unless otherwise specified. This charge and periodicity imply that the beam before IFEL micro-bunching has an easily accessible peak current of 380 A \cite{Robles2019}. The matched transverse spot size is calculated using $\sigma_x$ = $\sqrt{\beta_{eq} \epsilon_{\perp}}$. The total beam charge differs for each case based on the number of micro-bunches employed, and is specified with the plots.

\begin{table}
\caption{\label{tab:table1}Table of parameters for the simulation shown in Figure \ref{fig:10bunch}.}
\begin{ruledtabular}
\begin{tabular}{cc}
Parameter & Value \\
\hline
Plasma density, $n_0$ & $2.79$ $\times$ $10^{20}$ cm$^{-3}$\\
Beam charge, $Q_b$ & 3.2 pC\\   
Beam energy, $E_b$ & 10 GeV\\
Number of bunches, $m$ & 10\\
Bunch length, $\sigma_z$ & 100 nm\\
Bunch spot size, $\sigma_x$ & 13 nm\\
Normalized transverse emittance, $\gamma \epsilon_{\perp}$ & 50 nm-rad\\   
Plasma ion species & H$^+$
\end{tabular}
\end{ruledtabular}
\end{table}

\begin{figure}%
    \centering%
    \subfloat[]{{\includegraphics[width=1.65in]{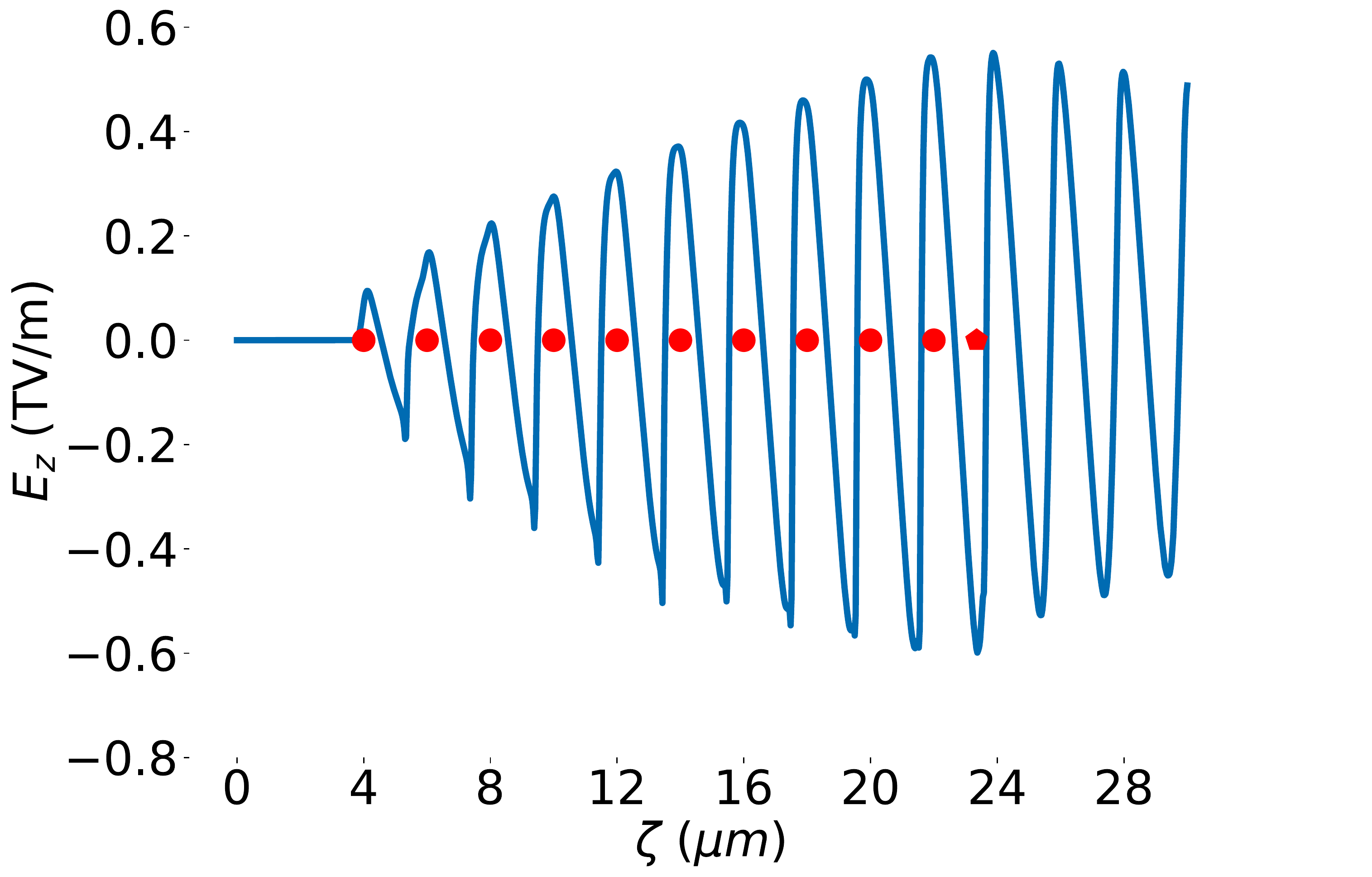} \label{fig:10-a}}}%
    \subfloat[]{{\includegraphics[width=1.65in]{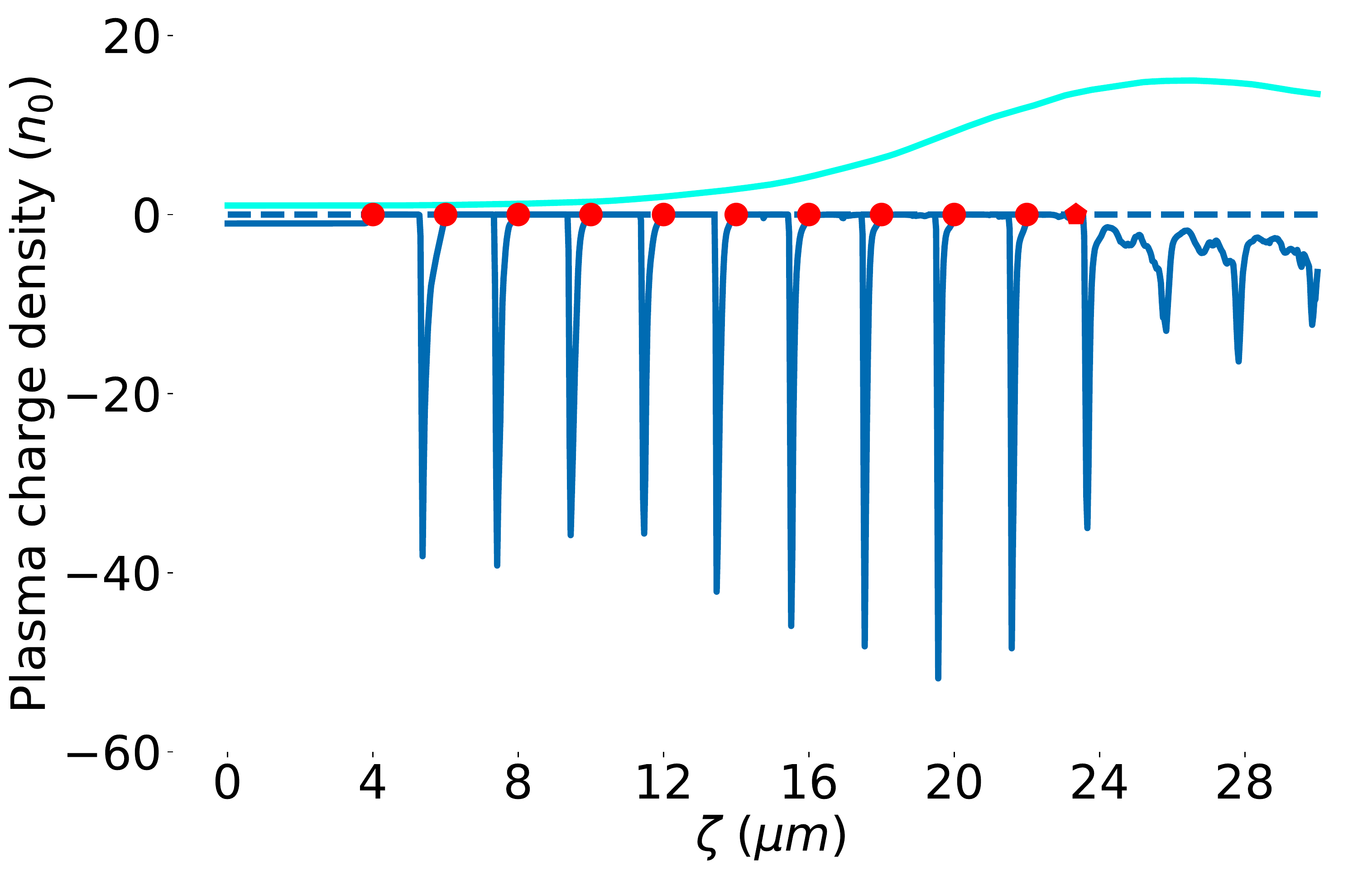} \label{fig:10-b}}}%
    \caption{The axial longitudinal electric field (a), plasma electron (dark blue) and ion (light blue) density (b) when using ten driver bunches (circle) (charge per driver bunch = 0.32 pC) and a witness bunch. Each bunch is separated by $\lambda_p$ and has an initial transverse spot size of 13 nm.}
    \label{fig:10}%
\end{figure}

\begin{figure}%
    \centering%
    \subfloat[]{{\includegraphics[width=1.65in,height=1.6in]{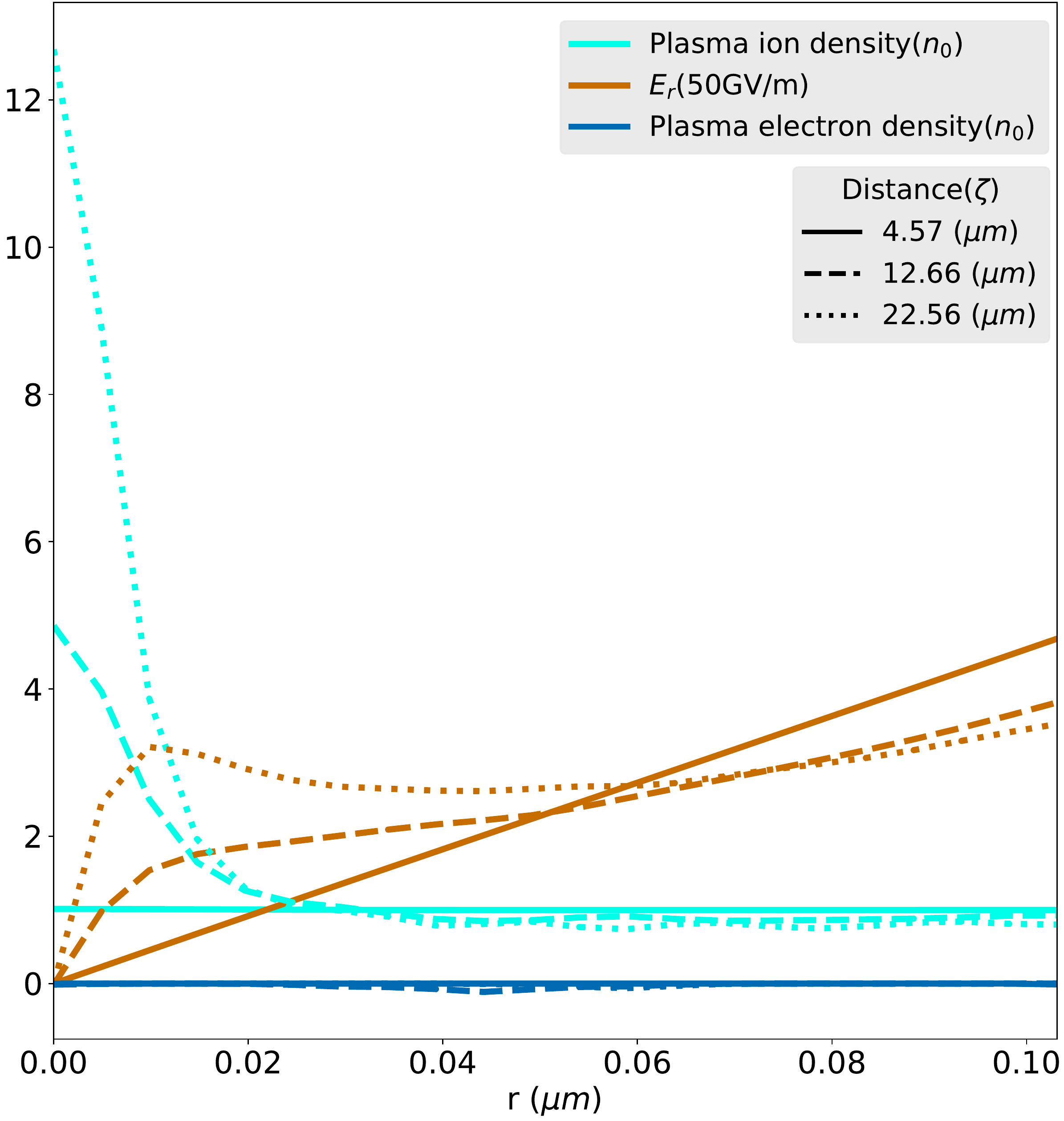} \label{fig:radial_in}}}%
    \subfloat[]{{\includegraphics[width=1.65in,height=1.6in]{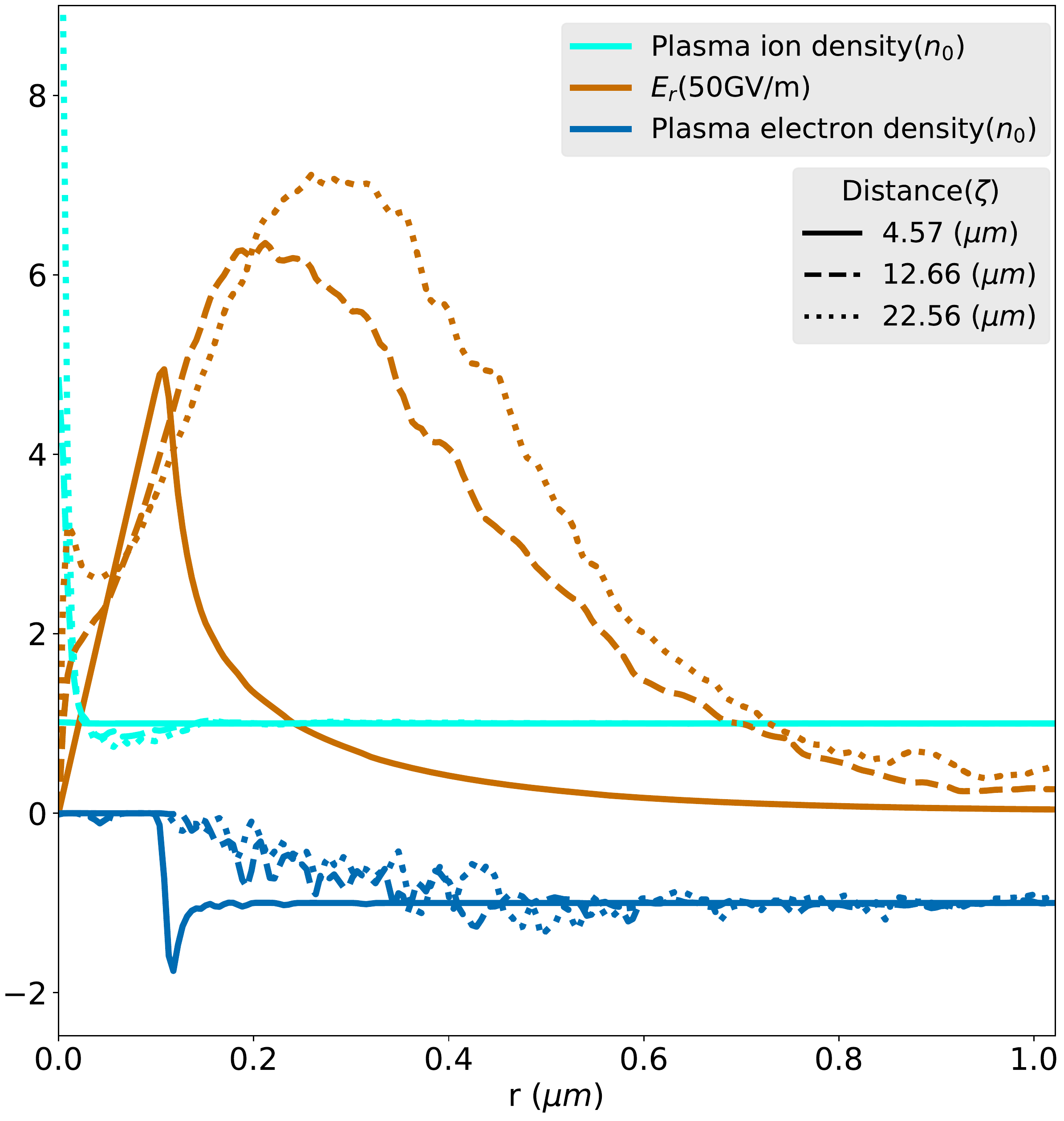} \label{fig:radial_out}}}%
    \caption{The radial fields, plasma electron and ion densities inside the bubble cavity at three different $\zeta$ positions. (a) shows the region near axis in greater detail.}%
    \label{fig:radial}
\end{figure}


A typical distribution of the beam charge density, plasma electron density, and longitudinal electric fields found in these studies are shown in Figure \ref{fig:10bunch}. The longitudinal electric fields increase after each micro-bunch passage, as shown in Figure \ref{fig:10-a}. The regions of very high plasma electron density at the ends of each bubble region become increasingly narrow after every subsequent micro-bunch due to nonlinear wave-breaking, as is shown in Figure \ref{fig:10-b}. 

Due to the very large electric fields involved, the typical approximation of static plasma ions is no longer valid and their distribution evolves over the course of the periodic beam-plasma interaction. The effect of ion motion on the plasma's ion and electron densities distributions, as well as associated radial electric fields, are shown in Figure \ref{fig:radial}. Given the presence of ion motion, each micro-bunch will experience notably different focusing fields leading to the observed ramping effect on the beam density profile. The on-axis ion density increase further augments the focusing gradient near the axis. It also can lead to nonlinear transverse fields which may contribute to emittance growth for distributions which extend well beyond the radially-localized ion density increase.

The system described above was simulated (see parameters of Table \ref{tab:table1}) for a time equal to $\frac{5000}{\omega_p}$, {\sl i.e.} a distance of about 1.59 mm. The resonant beam-plasma interaction remains stable for the entire duration of this simulation.  The maximum energy change observed was 950 MeV, corresponding to an average gradient of $\sim$0.6 TeV/m: a reasonable fraction of wave-breaking is achieved. 

\subsection{Variation with charge}

\begin{figure}%
    \centering
    \subfloat[]{{\includegraphics[width=1.65in]{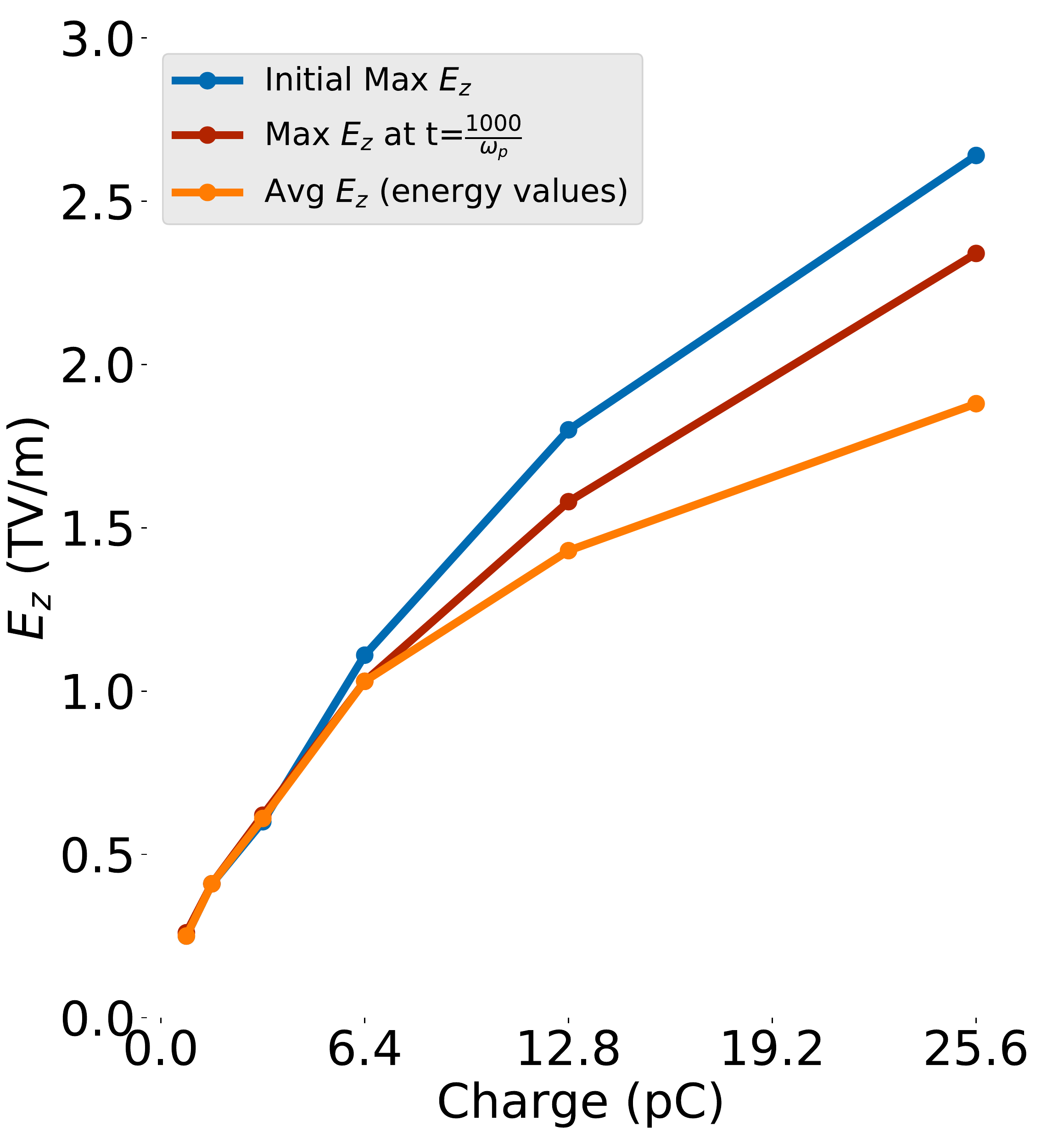} \label{fig:charge2}}}%
    \subfloat[]{{\includegraphics[width=1.65in]{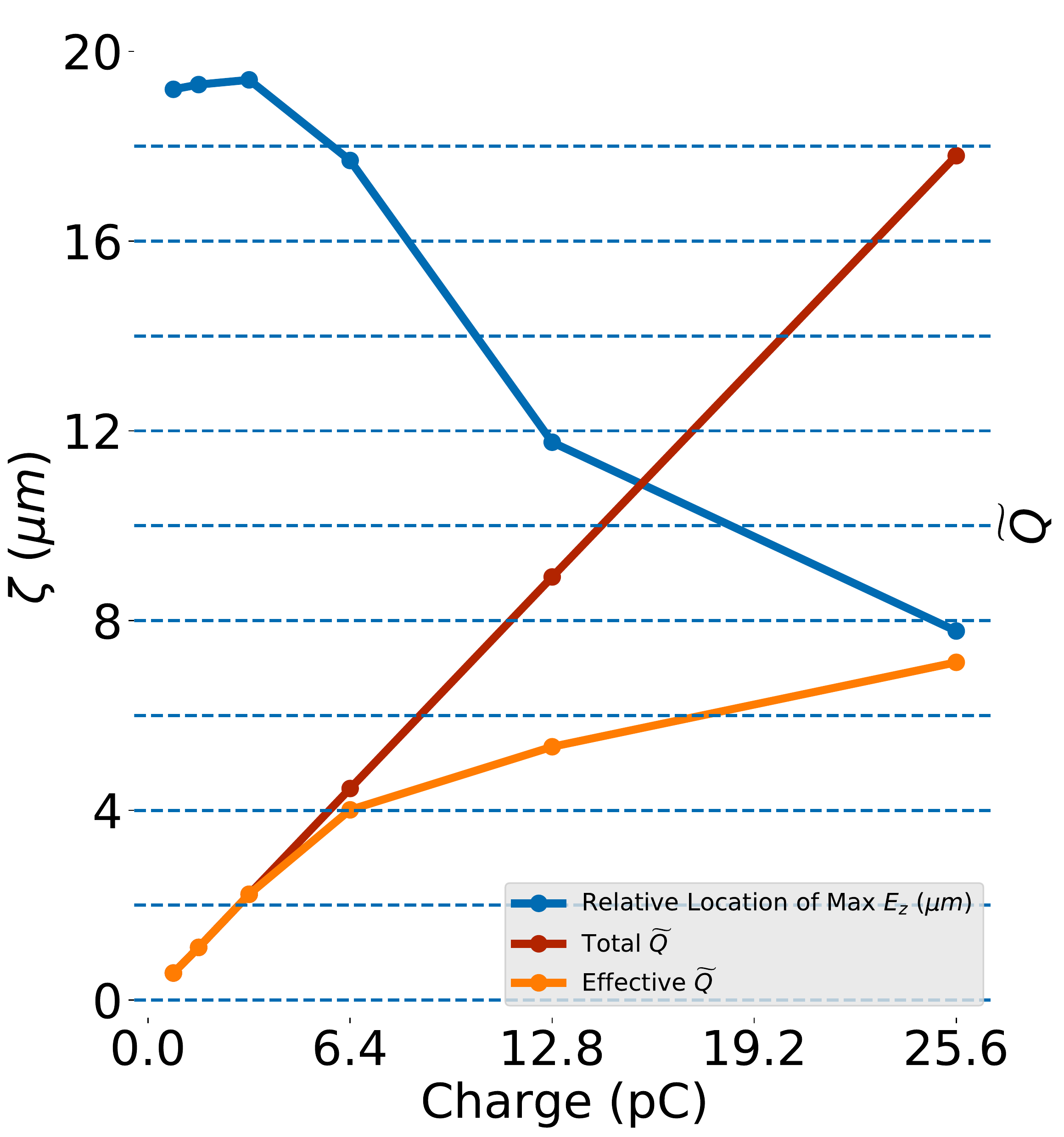} \label{fig:locationtilde}}}%
    \caption{The variation of initial, final and average maximum axial longitudinal electric fields with charge (a). Beam is completely bunched and emittance is scaled linearly with charge (3.2 pC $\rightarrow$ 50 nm-rad). The relative location of the peak longitudinal electric field, total and effective normalized charge density $\widetilde{Q}$ are also shown (b). The dashed lines correspond to the position of the bunches.}
    \label{fig:variation}%
\end{figure}

The relationship between beam charge and the maximum field gradient for this scenario with a ten micro-bunch train is now considered. The emittance in each of these cases was scaled linearly with the charge of the micro-bunch \cite{kim} while the longitudinal extent of each micro-bunch, $\sigma_z$, was kept constant. The simulations were run for $t = \frac{1000}{\omega_p}$ {\sl i.e.} a distance of about 0.32 mm, with the results shown in Figure \ref{fig:variation}. As the beam charge is increased, the maximum electric field increases, as the plasma perturbation is commensurately larger. Another feature observed is that the plasma wake profile tends to be less sinusoidal and more saw-tooth in form in the case of higher charges, a clear signature of the onset of wave nonlinearity. Due to such nonlinear effects, the saturation of the resonance response is achieved at an earlier point within the micro-bunch train. This reduces the effective charge that may be involved in resonant excitation, and the normalized charge up to this point can be termed as the effective $\widetilde{Q}$. There is a slightly weaker than linear correlation between the effective $\widetilde{Q}$ and the maximum longitudinal field. The fields from higher charge beams are also relatively diminished due to their higher assumed emittances. It can be observed  that the use of lower charge beams permits more micro-bunches to be involved in resonantly driving the wakefields.

The resonant wakefield is found to be sustained even for values of normalized charge density, $\widetilde{Q}$, as high as $2.23$ in the ten bunch case. As the effect of losing the resonant response is more notable for long micro-bunch trains; the value of $\widetilde{Q}$ may be yet higher if fewer micro-bunches are used. Permitted values of $\widetilde{Q}$ increase further with use of larger electron beam transverse spot size $\sigma_x$. This is because the plasma electron motion is dependent on the size of the perturbing field, {\sl i.e.} on the charge density of each bunch. Thus, in the multiple bunch case, even if the value of $\widetilde{Q}$ increases above unity, the wakefield frequency response remains nearly linear and can support resonant excitation for short trains. 

\begin{figure}%
    \centering
    \subfloat[]{{\includegraphics[width=1.65in]{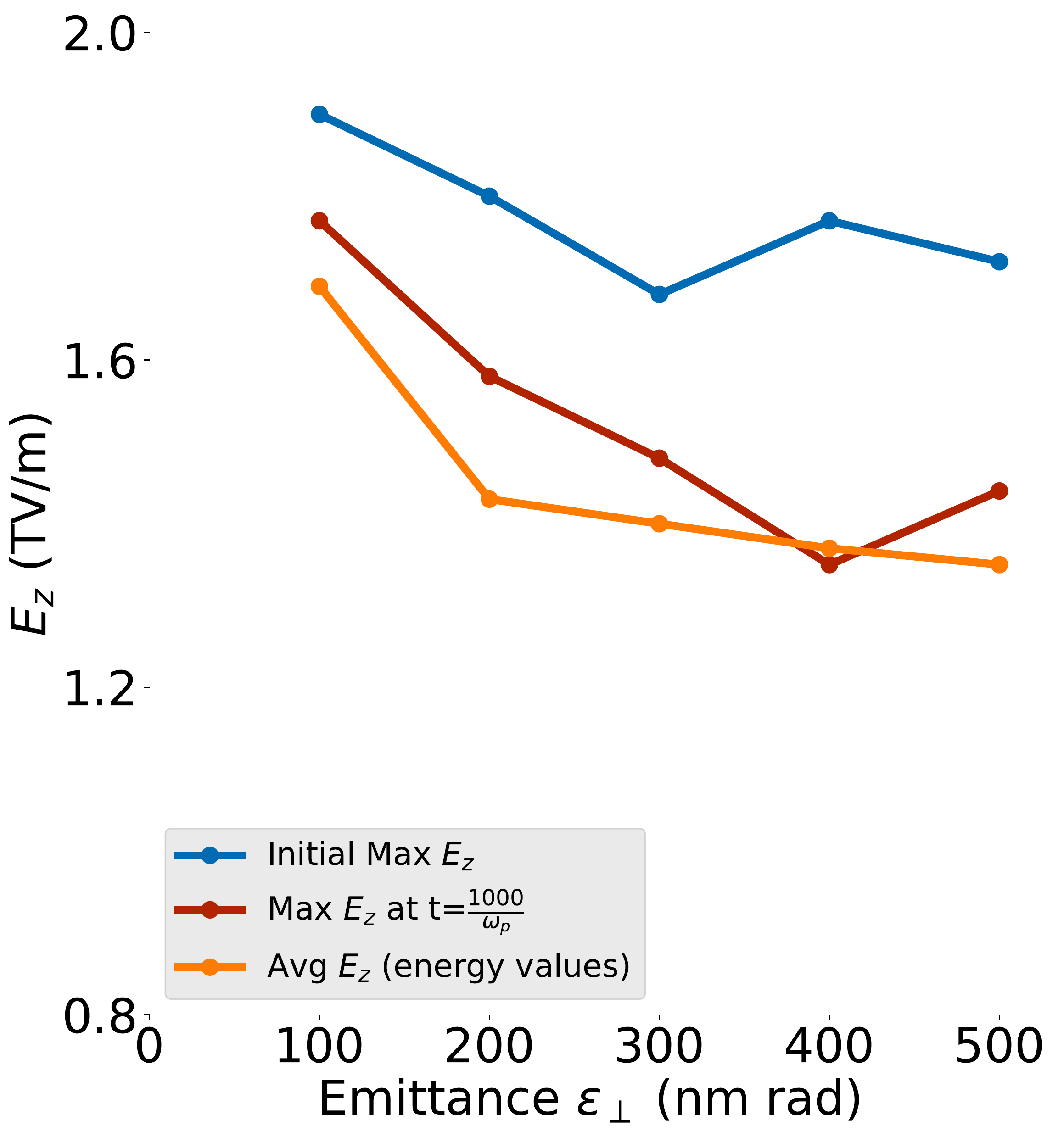} \label{fig:emittancefields}}}%
    \subfloat[]{{\includegraphics[width=1.65in]{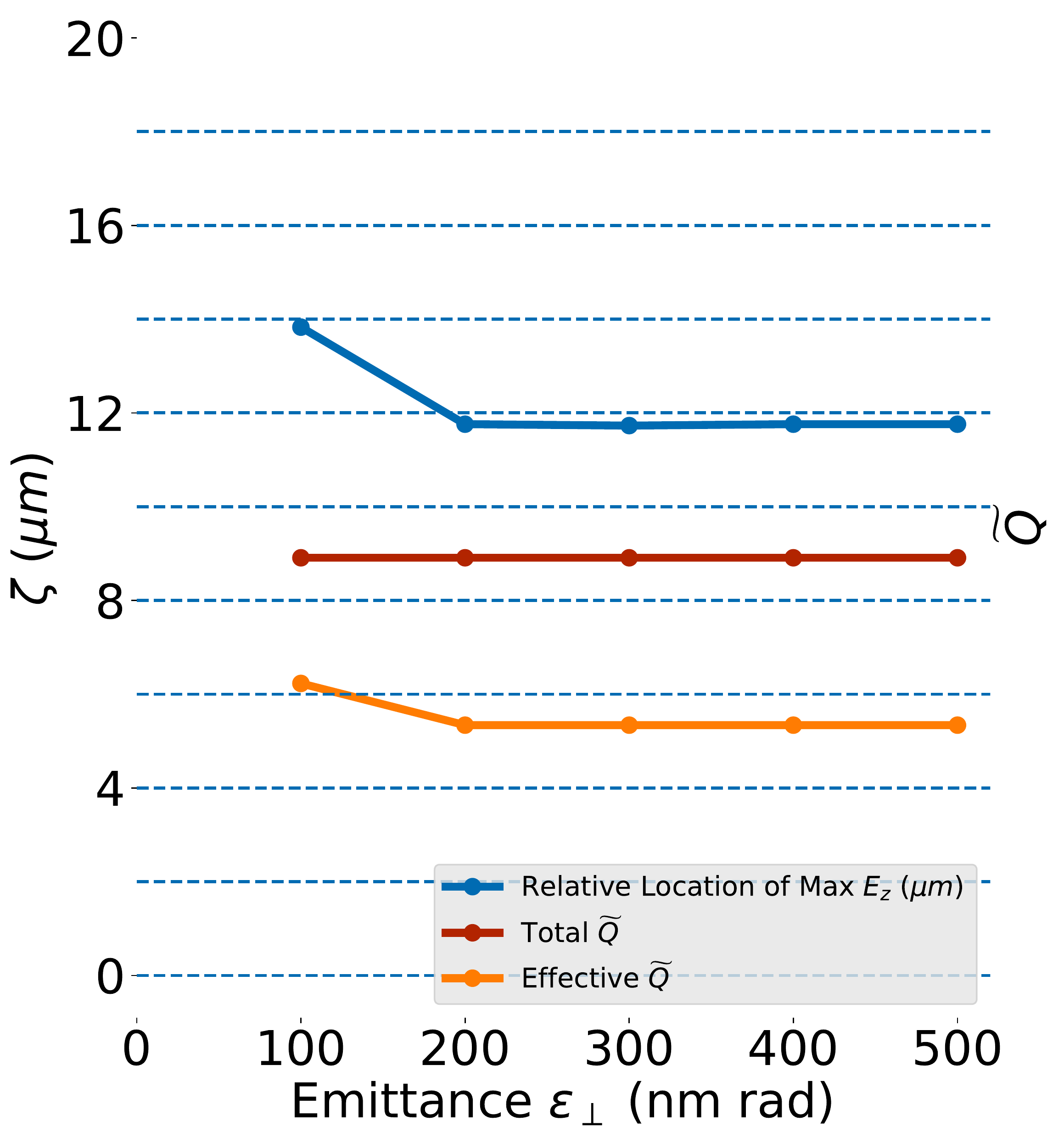} \label{fig:emittancetilde}}}%
    \caption{The variation of initial, final and average maximum axial longitudinal electric fields with emittance (a). Beam is completely bunched and beam charge is kept constant at 12.8 pC. The relative location of the peak longitudinal electric field, total and effective normalized charge density $\widetilde{Q}$ are plotted (b). The dashed lines correspond to the position of the bunches.}
    \label{fig:emittance}%
\end{figure}

\subsection{Variation with emittance}
To isolate its effect, the beam emittance in the simulations was varied while keeping the beam charge constant at 12.8 pC. The transverse beam spot size was matched with the plasma for each case. The variation of maximum longitudinal fields and average energy gradients with emittance are plotted in Figure \ref{fig:emittance}. The average field gradient tends to decrease with increasing emittance, as expected. This is mainly due to the increase in spot size, which decreases the peak beam density, resulting in a  weaker plasma wake.

\section{PARTIALLY BUNCHED SYSTEMS}

The density modulation given by simple application of IFEL micro-bunching is only approximately as described above.  Indeed, without using elaborate approaches \cite{Sudar2018}, the micro-bunching achieved will be partial, with non-negligible current in the pedestal between micro-bunches. Thus, practical considerations drive the need to understand the complications and possible advantages introduced in the resonant plasma wakefield system by imperfect, or partial, bunching.

\begin{figure}%
    \centering
    \subfloat[]{{\includegraphics[width=2.8in]{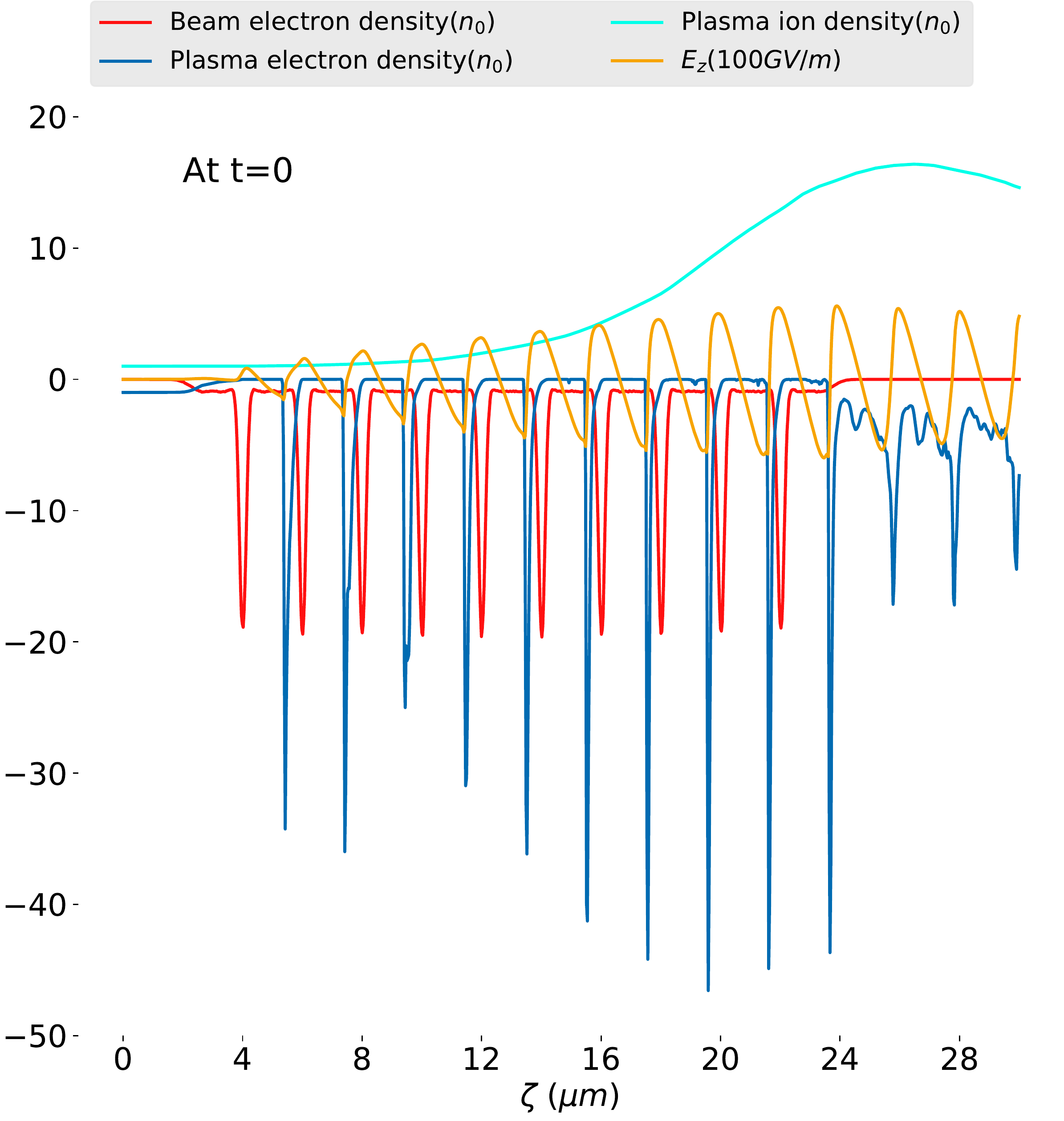} \label{fig:0}}}%
    \qquad
    \subfloat[]{{\includegraphics[width=2.8in]{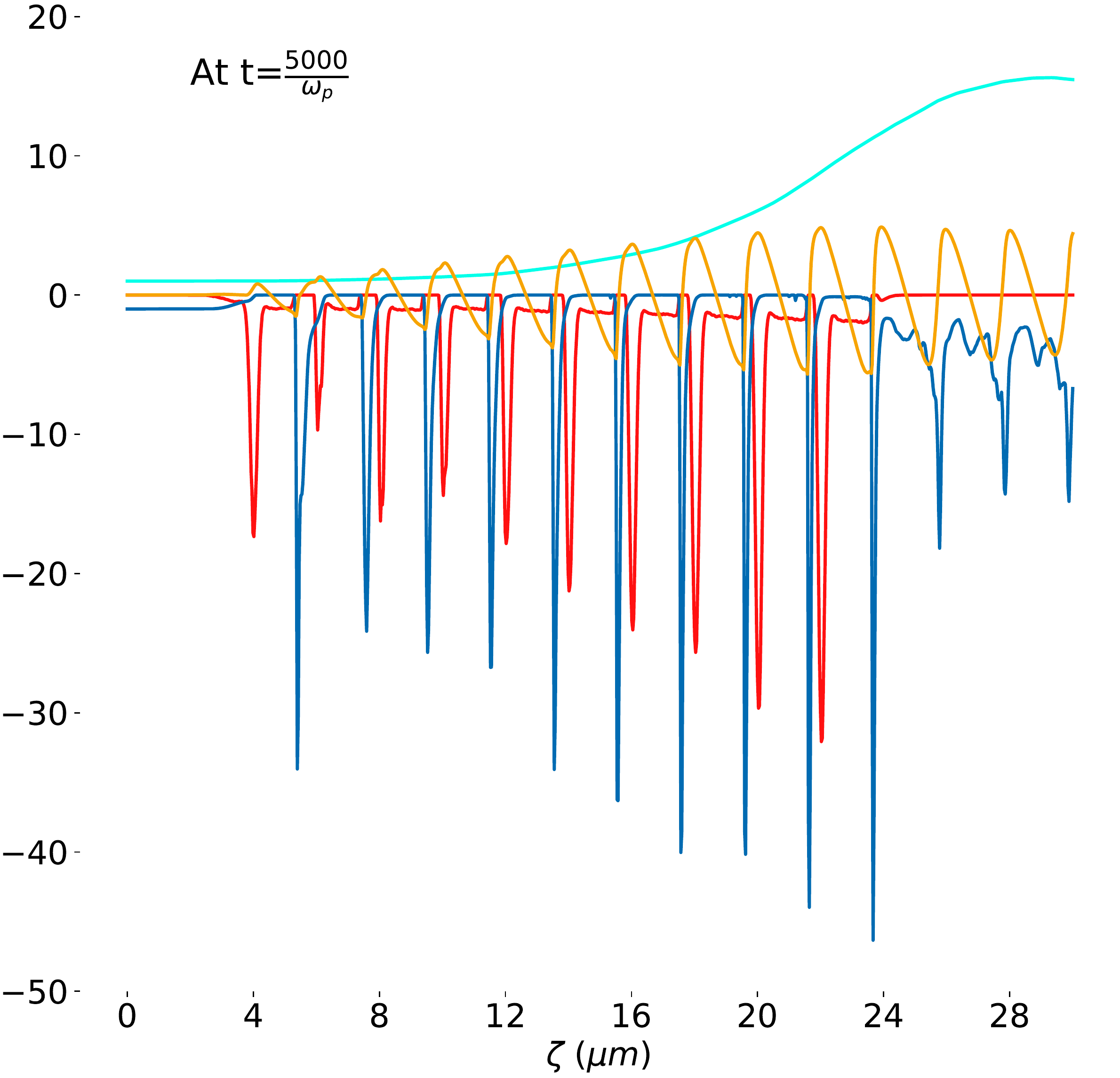} \label{fig:5000}}}%
    \caption{Beam charge density, plasma density, and longitudinal electric field of a partially bunched system (ratio between peak and flat region $\approx$ 20) initially (a) and after $t = \frac{5000}{\omega_p}$ (b).}
    \label{fig:time}%
\end{figure}

The partially bunched system described in Table \ref{tab:table2} was simulated for a time equal to $\frac{5000}{\omega_p}$, {\sl i.e} a distance of about 1.59 mm, with results shown in Figure \ref{fig:time}. The resonant PWFA interaction remains stable for this duration. The  beam  is uniformly bunched and focused initially, but assumes a ramped density structure as it propagates. This  ramping  is  due  to  stronger  focusing  of the  trailing  bunches, as before,  by  the  higher  on-axis  ion  density after the onset of ion motion. The maximum energy change was observed to be 923 MeV corresponding to an average accelerating gradient of 0.58 TeV/m. The tail of the beam, {\sl i.e.} a part of the unbunched beam component, was employed as the witness beam to enable this calculation of energy change.

\begin{table}
\caption{\label{tab:table2}Table of parameters for the simulation shown in Figure \ref{fig:time}}
\begin{ruledtabular}
\begin{tabular}{cc}
Parameter &Value\\
\hline
Plasma density, $n_0$ & $2.79$ $\times$ $10^{20}$ cm$^{-3}$\\
Beam charge, $Q_b$ & 4.21 pC\\
Beam energy, $E_b$ & 10 GeV\\
Number of bunches, $m$ & 10\\
Charge distribution & From IFEL ($B$=0.71)\\
Beam spot size, $\sigma_x$ & 14.6 nm\\
Normalized transverse emittance, $\gamma \epsilon_{\perp}$ & 66 nm-rad\\
Plasma ion species & H$^+$
\end{tabular}
\end{ruledtabular}
\end{table}

\subsection{Variation with bunching factor}

\begin{figure}
    \centering
    \includegraphics[width=3.3in]{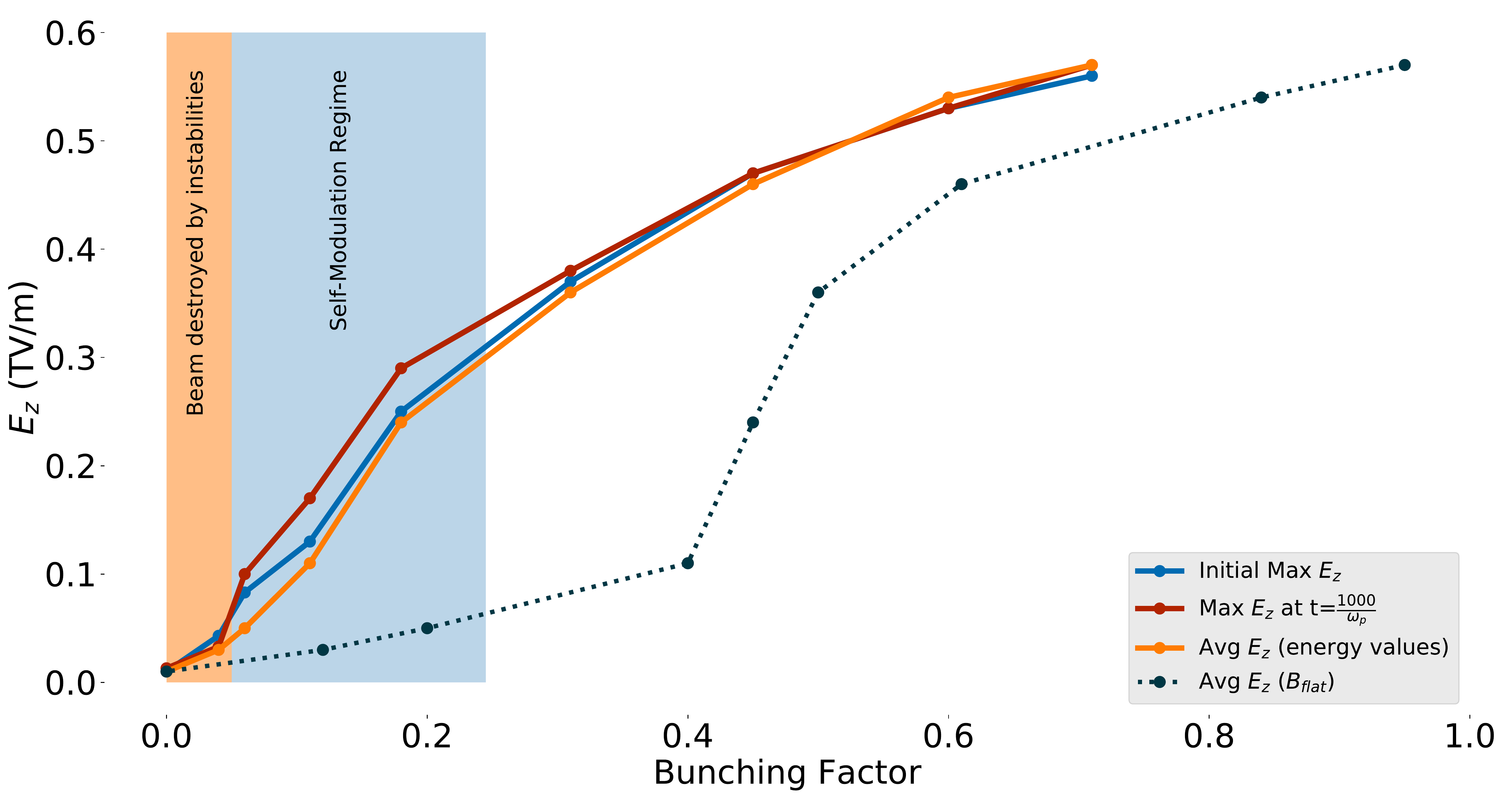}
    \caption{The variation of initial, final and average maximum axial longitudinal electric field with bunching factors ($B$, $B_\mathrm{flat}$).}
    \label{fig:bunch}
\end{figure}

The relationship between the beam and plasma responses and the bunching factor is important to understand, due both the practical difficulty of approaching full bunching, as well as possible advantages due to incomplete bunching. A common figure of merit for micro-bunching is the \emph{bunching factor}, defined as $B =\sum{\mathrm{e}^{i\theta_{i}}}/N_b$, where $\theta_i$ are the particle longitudinal phases at a selected wave number \cite{hhifel}. This quantity is equal to zero in the case of a uniformly distributed beam and tends to unity in the case of a beam perfectly micro-bunched at the bunching wavelength. However, this does not give the necessary information concerning the locally nonlinear beam-plasma interaction. We are, in this regard, most interested in the ratio of the peak beam current in the micro-bunches to the current between them, where there is a nearly flat pedestal.  It is therefore useful to introduce a different factor, $B_\mathrm{flat}$, which only considers the portion of the current above this pedestal. In the simulations involving partially bunched systems described above,  $B_\mathrm{flat} = 0.93$, notably larger than the value of $B$ from the standard definition. The variation of the longitudinal electric fields with bunching factor $B$ and the corresponding value of $B_\mathrm{flat}$ is shown in Figure \ref{fig:bunch} along with different regimes of operation detailed below. Generally, the transverse and longitudinal fields are larger at higher bunching factors due to the enhancement of the perturbing beam density and the strengthening of frequency content of the micro-bunch train at the resonant frequency. Additionally, with stronger bunching the phase slippage of the wake (discussed in greater detail below) is reduced, preserving the stable interaction over greater distances.

\begin{figure}%
    \centering
    \includegraphics[width=3.3in]{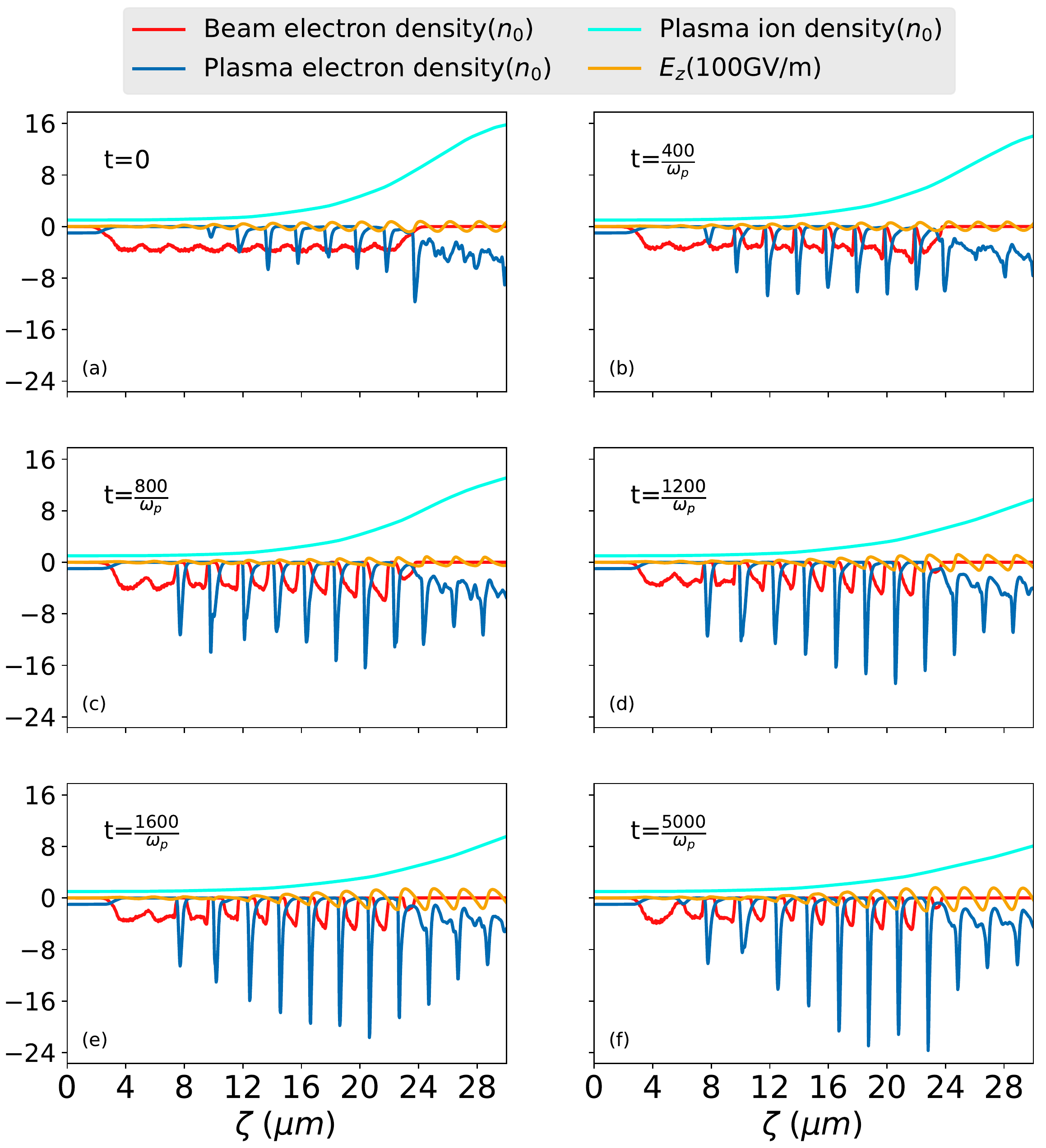}%
    \caption{Self modulation regime ($B$=0.06); The axial longitudinal electric fields (yellow), beam densities (red) and plasma densities (blue) are shown. Initially, plasma electrons are unable to form voids in the beam \hyperref[fig:self-modulation2]{(b)}. As the beam propagates, the self-modulation instability forms these voids \hyperref[fig:self-modulation2]{(e)} and resonance is maintained for a long duration \hyperref[fig:self-modulation2]{(f)}. Notice, the amplification of the electric fields and the backward shift of the plasma wake with time.}
    \label{fig:self-modulation2}%
\end{figure}

Partial micro-bunching may be viewed as similar to seeded self-modulation \cite{seeding} as it allows one mode to be preferentially excited in the beam-plasma system. If this incomplete bunching is sufficiently high ($B \gtrsim 0.05$), the stable self-modulating mode grows, continuing to modulate the beam profile and increasing the bunching factor. Specifically, early in the beam-plasma interaction, the plasma electrons are completely ejected near the head of the beam (See Figure \hyperref[fig:self-modulation2]{8a}), leaving an electron-rarefied column several $\lambda_p$ in length. As the interaction proceeds though, plasma electrons are attracted back towards the axis by the ion column's radial field. This in turn causes the defocusing of beam electrons far from the nominal bunching phase, increasing the beam's bunching factor by ejecting these electrons radially outward into a halo distribution, and reducing the on-axis ion density in the process. This feedback loop, termed the self-modulation instability (SMI), is convective and is therefore stronger further from the head of the beam \cite{proton-self}. The growth of this instability is illustrated in Figure \ref{fig:self-modulation2}; the SMI-induced effects ultimately approach a quasi-steady state where the beam and plasma profiles' evolution slows down over the duration of the simulation, $5000/\omega_p$. 

\begin{figure}%
    \centering
    \subfloat[]{{\includegraphics[width=2.8in,height=1.3in]{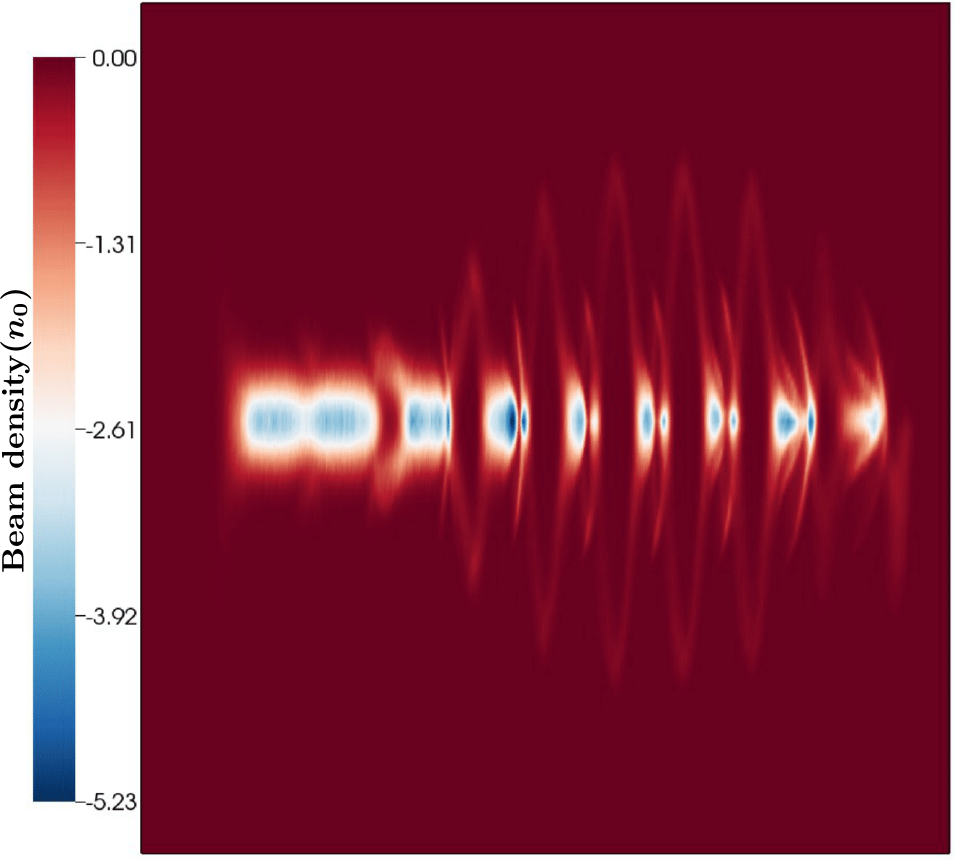} \label{fig:decoherence}}}%
    \qquad
    \subfloat[]{{\includegraphics[width=2.8in,height=1.3in]{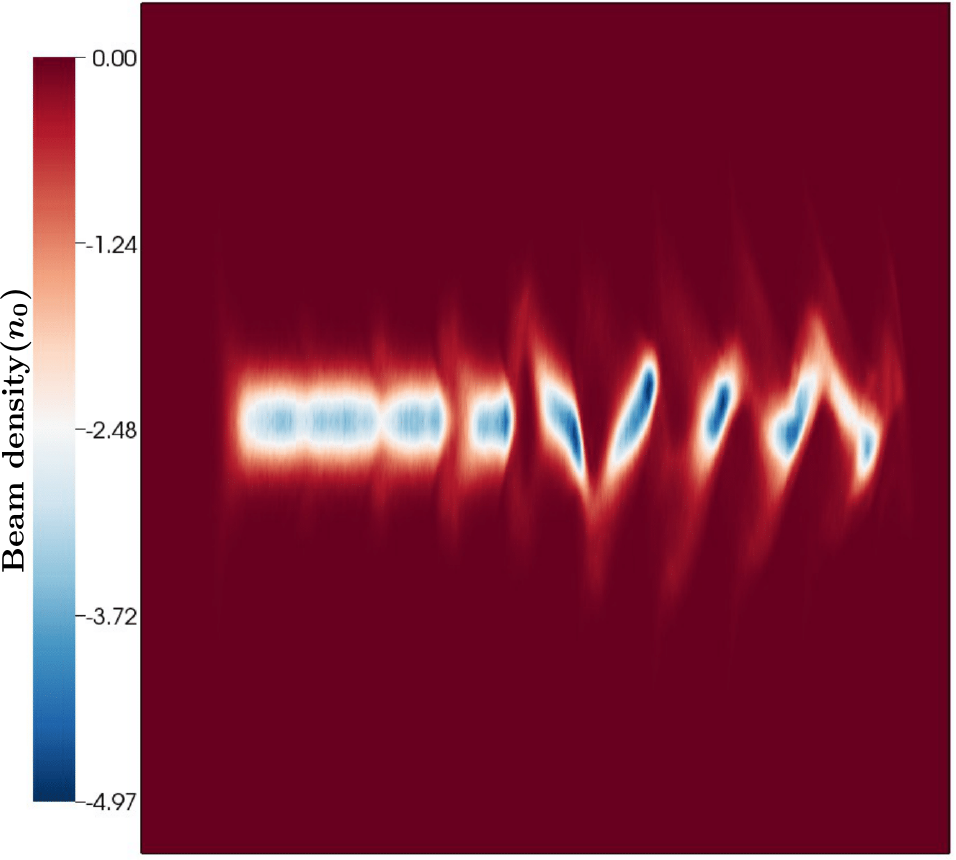} \label{fig:hose}}}%
    \caption{Observed beam destruction due to phase shifting of the plasma wake at $B$=0.04 (a) and hosing instability at $B$=0.02 (b).}
    \label{fig:breakup}%
\end{figure}

However, at lower bunching factors, other instabilities may destroy the beam before SMI can produce a stable beam profile \cite{proton-self}. For beams with bunching factors slightly below what is determined to be the stable SMI threshold ($0.03 \lesssim B \lesssim 0.05$), the preferred modulation mode does not dominate as quickly, permitting other modes to develop. This effect has been previously described \cite{lotov}: defocusing regions slip backwards along the beam, ultimately ejecting all beam electrons not contained within the leading bunch. A snapshot of this process is shown in Figure \ref{fig:breakup}(a) for a beam with $B = 0.04$.

At very low bunching factors ($B \lesssim 0.03$) the beam is effectively a long, uniform beam. Since the growth rates of SMI and the \textit{hosing instability} are comparable \cite{coupled-instable}, with only weak seeding of SMI, the hosing instability also manifests and contributes to the prompt destruction of the beam. Both SMI and the hosing instability are evident in Figure \ref{fig:breakup}(b) for a beam having initial bunching factor $B = 0.02$.

\subsection{Variation with charge, partially bunched case}

\begin{figure}%
    \centering
    \subfloat[]{{\includegraphics[width=1.65in]{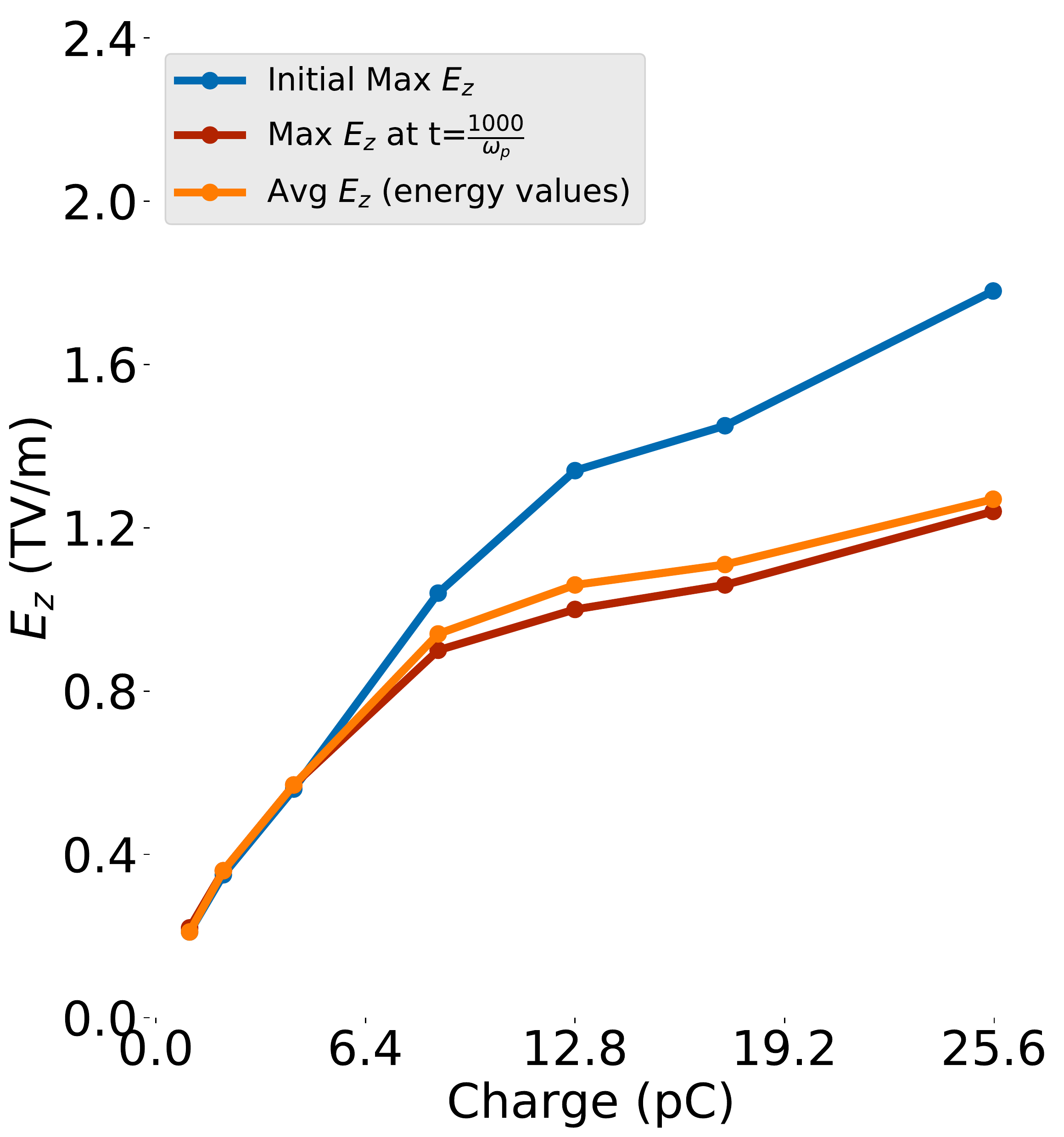} \label{fig:partialcharge}}}%
    \subfloat[]{{\includegraphics[width=1.65in]{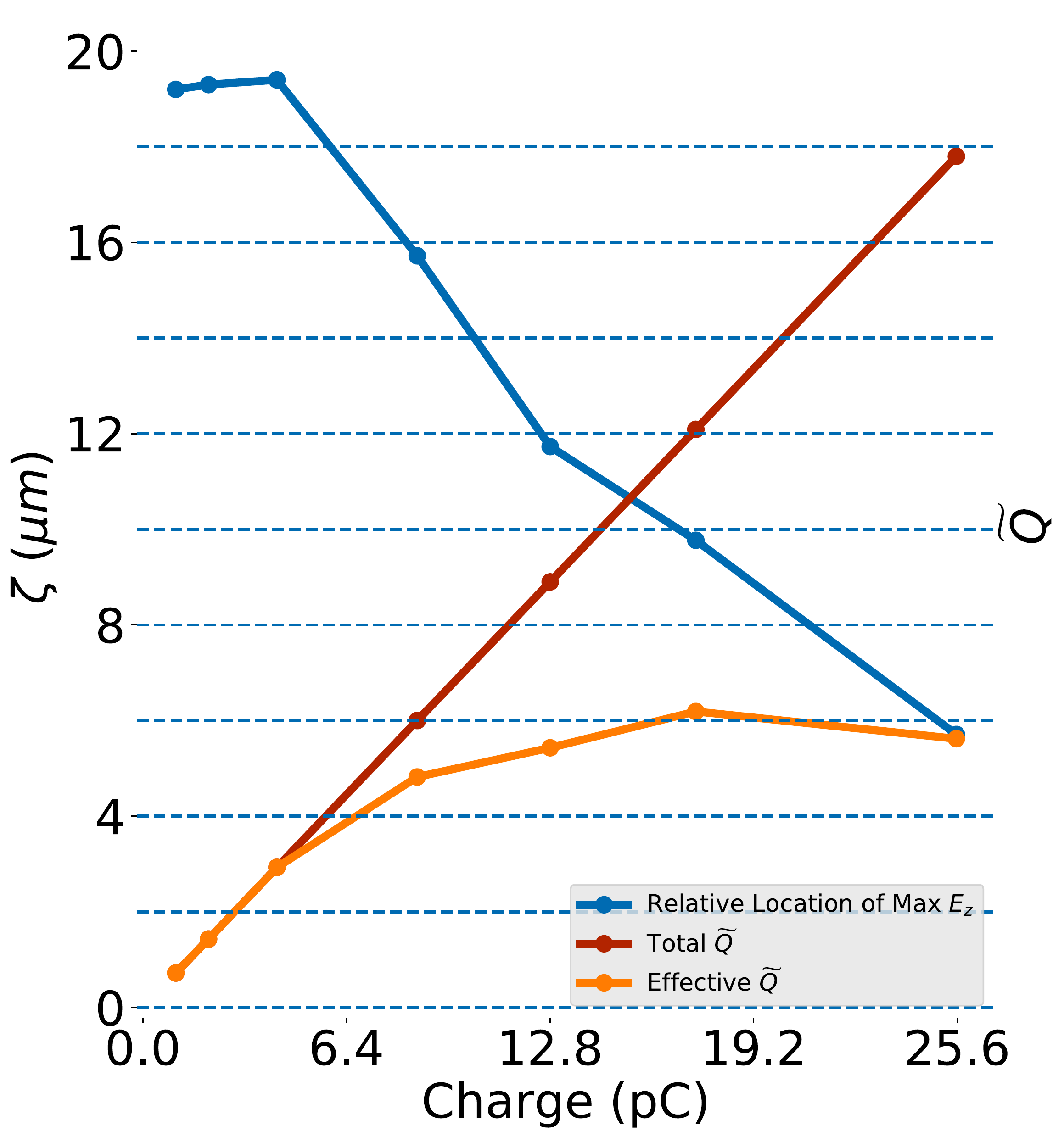} \label{fig:partiallocationtilde}}}%
    \caption{The variation of initial, final and average maximum axial longitudinal electric fields with charge are plotted in (a). Beam is partially bunched ($B$ $\approx$ 0.71) and the ratio between the peak and flat region $\approx$ 20. Emittance is scaled linearly with charge (3.2 pC $\rightarrow$ 50 nm-rad). The relative location of the peak longitudinal electric field, total and effective normalized charge density $\widetilde{Q}$ are plotted in Figure  (b). The dashed lines correspond to the position of the bunches.}
    \label{fig:partialvariation}%
\end{figure}

Here we examine the dependence of the longitudinal fields on beam charge in the context of a partially bunched beam. The bunching factor was held constant at $B=0.71$ (corresponding to $B_\mathrm{flat}$=0.93), keeping the ratio between the peak and flat region constant at about 20. As before, the beam emittance was scaled linearly with charge. The variation of the fields are shown in Figure \ref{fig:partialvariation}. The effects of higher charge on the fields are similar to the fully bunched case (Figure \ref{fig:variation}), but the slope of the final field with respect to charge is shallower due to the effect of the flat-current pedestal. The voids formed in the beam due to the plasma electron's return to the axis are less pronounced at higher charges due to the steeper longitudinal dependence of the plasma wake. The value of effective $\widetilde{Q}$ increases with larger charge micro-bunches and, as before, the resonance eventually saturates. However, this saturation occurs at lower $\widetilde{Q}$ than in the completely bunched case. 

\section{OTHER INSTABILITIES IN RESONANT EXCITATION}

The resonant beam-plasma system has inter-dependencies between the electron micro-bunches and the plasma response, leading to both stable interactions as well as unstable scenarios. The dominant instabilities arise from the loss of charge of the bunches and the phase shifting of the plasma wake, both of which can manifest in high $B$ cases. We observe in Figure \ref{fig:10bunch} the additive increase in the plasma bubble dimensions and plasma electron density after each bunch. The first instability, namely charge loss, results from the ejection of parts of the drive bunches that are near the regions of high plasma electron density, reducing both the beam charge and the strength of $E_z$. The second instability, phase shifting, begins due to head erosion of the leading bunch. As it loses charge, there is a decrease in the plasma electron perturbation causing a backwards shift of the wake (negative $\zeta$). The second bunch begins to experience a defocusing force, reducing its charge and propagating the instability back along the rest of the bunch train. This same instability also manifests if the plasma wavelength is not well matched with the bunch spacing. For a well-matched bunch train, this instability can be mitigated by the use of a lower emittance beam, and by employing a ``pilot" focusing mechanism for the first bunch, such as a preceding intense laser pulse or a leading, tailored component of the electron beam. 

Incomplete bunching can produce a nearly flat region of pedestal current at the front of the bunch train to serve as the pilot beam. This section of the beam produces some ion channel focusing fields which help slow the erosion of the more sensitive, leading micro-bunch, helping diminish the growth rate of the phase shifting instability. This scenario has been studied through simulations with an example shown in Figures \ref{fig:0} and \ref{fig:5000} which includes the onset of head erosion effects. The resonant system was also simulated without the pilot component. Although the initial fields were higher in the large-$B$ case, they decreased much more quickly due to head erosion. For comparison, the case without the pilot was simulated for the same, short distance (1.59 mm) and exhibited a maximum energy gain of 905 MeV, corresponding to an average gradient of 569 GeV/m, marginally lower than the pilot-free case. The possible improvements obtained from the pilot section must be weighed against the instabilities present in partially bunched systems. Additionally, if the plasma is to be formed by beam-based ionization, the pilot may not be sufficient to completely ionize the plasma. Although a partially ionized plasma will still exert beneficial focusing on the leading micro-bunch, it is not as effective, and supplementation with another ionization method, such as a leading laser pulse, may be desirable.


\section{WITNESS BEAM INJECTION}

The dimensions of the plasma bubble created by optically micro-bunched beams are notably smaller than the accelerating region currently found in PWFA experiments. If the beam is not fully micro-bunched by the IFEL process, the electrons in the flat pedestal of the current profile may be trapped by the large accelerating wakefields, forming a self-injected witness train. A smaller bunching factor leaves a larger number beam electrons available for injection, but a relatively small fraction would be trapped at accelerating phases. Additionally, since electrons in the pedestal are relatively diffuse, the energy spread of a self-injected witness would be relatively large. Alternatively, laser induced ionization injection, such as the Trojan Horse technique \cite{laser,Deng2019}, might be used to produce a brighter witness. This method is challenging in a variety of ways, however. First, the ionized region is not small compared to the bubble. Further,  in the QNL regime, the wakefield amplitude may be notably below wave-breaking (here we see $<0.4 \times E_\mathrm{WB}$), which implies that trapping may be difficult to achieve. 

\section{BEAM-INDUCED FIELD IONIZATION}

\begin{figure}%
    \centering
    \subfloat[]{{\includegraphics[width=2.8in,height=1.2in]{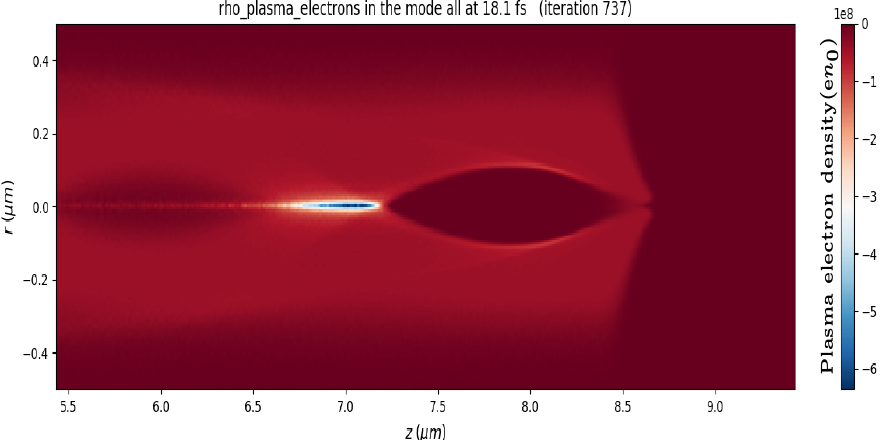} \label{fig:electrons}}}%
    \qquad
    \subfloat[]{{\includegraphics[width=2.8in,height=1.2in]{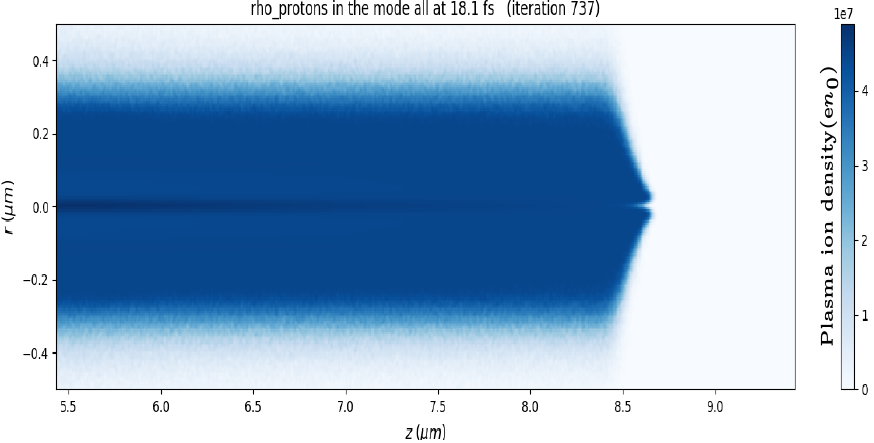} \label{fig:protons}}}%
    \caption{The plasma electrons (a) and protons (b) created by the field ionization of hydrogen gas by a single electron bunch positioned at $z = 8.44$ $\mu$m (charge per bunch = 0.32 pC, $\sigma_x$ = 13 nm, $\sigma_z$ = 110 nm).}
    \label{fig:plasma}%
\end{figure}

Beam-based ionization is a critical topic for a very high field scenario of resonant PWFA excitation using optically micro-bunched beams. This is owed to the fact that space-charge fields near these extremely strongly focused, very high brightness electron micro-bunches are near the TV/m level. This field is large enough to ionize the plasma in a few femtoseconds \cite{tev}. Such fast ionization has been confirmed in simulations using FBPIC \cite{fbpic}, where the plasma is indeed formed within a few femtoseconds of beam passage. The plasma wake formed in this case is shown in Figure \ref{fig:plasma}. The probability rate of ionization \cite{ionization} implemented in FBPIC follows the Ammosov-Delone-Krainov (ADK) tunneling ionization model \cite{adk}. The gas used in these simulations was hydrogen, in order to avoid multiple ionization. The first micro-bunch of the partially bunched beam was used in this simulation to demonstrate the efficacy of the ionization process. 

It should be noted  that the beam fields may cause ionization of atomic species in the barrier suppression ionization (BSI) regime \cite{bsi} wherein electrons are classically permitted to escape from the nuclear potential, and tunneling is not a relevant concept. This effect allows the plasma wake to be formed even more quickly than indicated by tunneling models. This fast ionization process may obviate the need to pre-ionize the gas. However, there are two major obstacles to reliance on BSI. First, the BSI regime is still not theoretically well-understood since it requires a non-perturbative approach. Second, for the leading edge of the beam, the threshold for full-ionization via BSI may not be reached due to head erosion, and a mix of BSI and tunneling will be present in self-ionized scenarios \cite{head_self}. To avoid this, pre-ionization may still be desirable. 

\section{DISCUSSION AND OUTLOOK FOR FACET-II EXPERIMENTS}

The experimental realization of TV/m plasma wakefields through single bunch, low emittance beam excitation has been discussed in some detail in Ref. \cite{tev}. Here we have presented and discussed in some detail an alternative path to achieving TV/m wakes, through a quasi-nonlinear resonant excitation mechanism. A key advantage here is shared with the application of  very low emittance beams to driving an x-ray FEL -- the final compression of the beam (which yields resonance in a very high density plasma in the QNL regime) is performed via IFEL. This permits obtaining of very high peak current while avoiding many of the deleterious effects of conventional compression. Experimental preparation in both QNL PWFA and x-ray FEL cases begins with the creation of the beam, and in this regard it is noted that since the time of the experiments reported in Ref. \cite{ding}, and the initial analysis of \cite{tev}, that significant improvements in the approach to obtaining higher brightness beams have been introduced, \textit{e.g.} Ref. \cite{topgunRosenzweig}. Indeed, in \cite{topgunRosenzweig}, an analysis of IFEL-induced micro-bunching for ultra-high brightness, 10 GeV-class beams (as would be employed for FACET-II experiments \cite{joshi_facet}) has already been performed. The efficacy of IFEL micro-bunching has been validated by the successful results of the XLEAP experiment at SLAC \cite{exleap}, with the achievement of high brightness micro-bunches verified through the generation of attosecond x-ray FEL pulses. To enable the experimental scenarios analyzed here, one must introduce, in addition to a high brightness electron source, a laser modulation and bunching system as employed in XLEAP.

After creation, acceleration, compression and micro-bunching of the electron beam at FACET-II, the beam must be focused to very small spot sizes, sub-mm at these high plasma densities \cite{tev}. This can be accomplished using very high gradient (700 T/m, or higher) permanent magnet quadrupoles that are 10's of cm in length, and tuned via changing their relative longitudinal positions \cite{magnet}. Such a focusing system will also be needed for the FACET-II experiment E-314 on ion motion, with its attendant search for formation of ion-beam electron focused Bennett-form equilibria \cite{bennett}. Alternatively, one may use an underdense plasma lens for the final focusing element, as already proposed for FACET-II \cite{facetlens}, with implementation now being initiated. 

\begin{figure}
    \centering
    \includegraphics[width=3.3in]{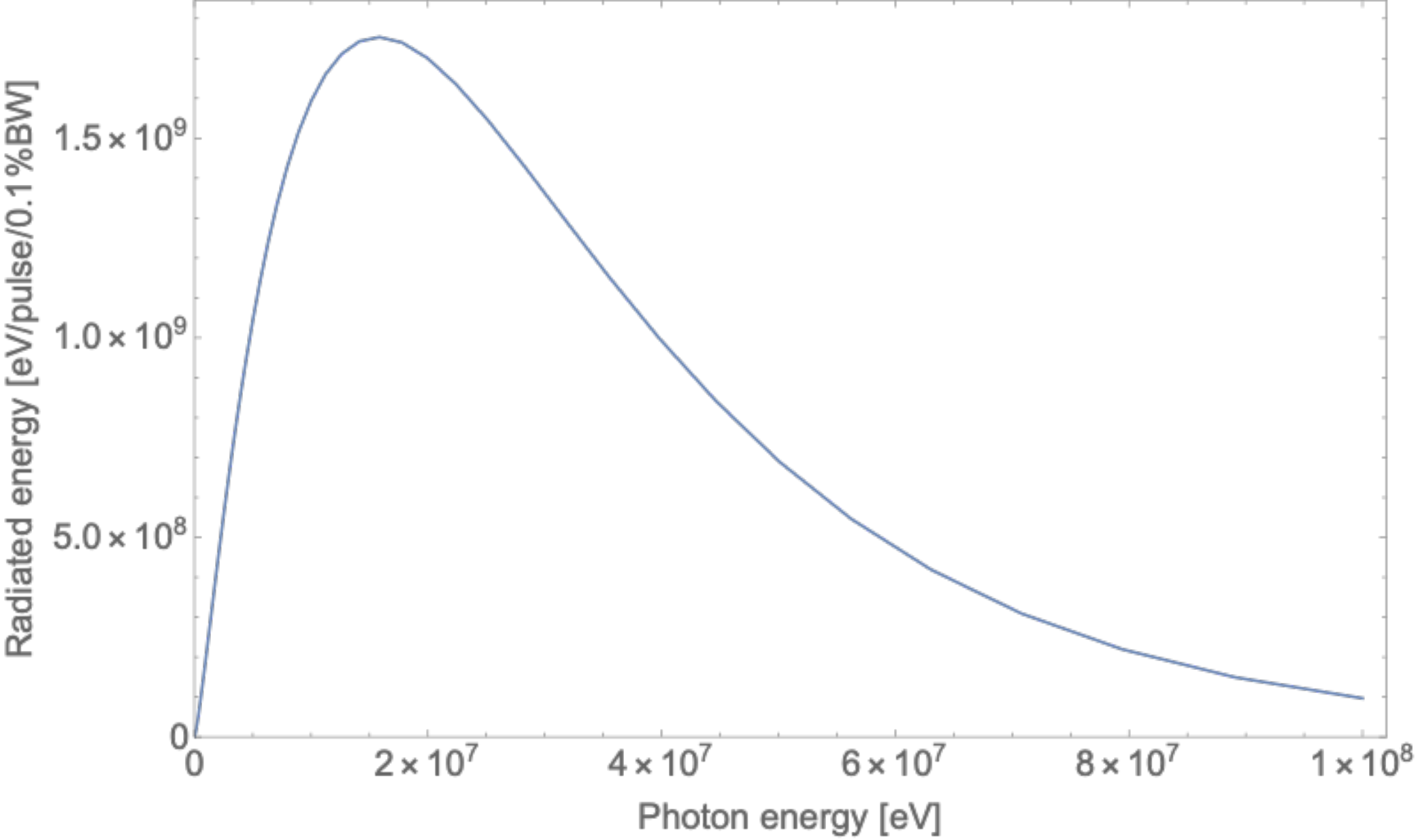}
    \caption{Betatron radiation emitted by a single electron bunch (charge=0.32 pC, $\sigma_x$ = 13 nm, $\sigma_z$ = 100 nm) as it propagates a distance of 1.6 mm with $n_i$=$n_0$; ion collapse effects are not included.}
    \label{fig:rad}
\end{figure}

The beam size at final focus may be deduced from appearance intensity \cite{tev}, and ionization yield in the gas, which is to be supplied in FACET-II experiments by nozzle jets to achieve several mm of multi-atmosphere pressure hydrogen. Alternatively, with an optically micro-bunched beam, one may use coherent diffraction imaging (CDI) based methods to reconstruct the sub-optical beam profiles. This is due to the tight bunching employed, which can give high harmonics of the bunching period in coherent emission processes (\textit{e.g.} edge radiation, plasma-based transition radiation). This short wavelength light can permit coherent imaging reconstruction, thus extending the IFEL-based CDI beam profile measurements reported in Ref. \cite{AgoCDI} to smaller spot sizes.  Once established, the beam-plasma interaction can be interrogated by the measurement of the betatron radiation spectrum. An example of such a spectrum, for a case without hosing or ion collapse, is shown in Fig. \ref{fig:rad}. The radiation spectrum in this preliminary analysis is quite hard, extending to beyond 100 MeV, due to the very high plasma density and concomitant focusing strength. Studies of the changes to this spectrum due to instabilities and attendant larger amplitude betatron motion are now under way. 

Many diagnostic systems needed for characterizing the beam will be available at FACET-II \cite{facetlens,FACETII}. These include: the betatron radiation spectrum via a Compton/pair spectrometer, as described in Ref. \cite{CPT}; the downstream beam imaging systems to determine phase space dilution of accelerated beams in this case \cite{ramp}; and momentum-resolving spectrometers. It should be noted that this experiment, along with that aimed at demonstrating the development of ion-electron beam Bennett equilibria, should be the first investigations to unequivocally access a regime of non-trivial ion motion, a critically important effect in linear collider applications of PWFA \cite{ioncollapse,Weiming2017}. In addition to observable focal effects on the beam, and related betatron radiation signatures, it is proposed to instrument the interaction region with a keV-range ion retarding energy analyzer.  

The existence of independent witness beams at FACET-II are a key feature of the new facility, which should be exploited for injection studies in this system. Injection may also be accomplished by detuning the energy of the beam tail to prevent effective micro-bunching in this region. This would permit the loading of electrons at a full range of initial phases, leading to a large fraction of witness beam electrons being captured and accelerated. 

The issues to be explored in this proposed FACET-II experimental program are myriad. The effects of the bunching factor are of particular interest, as they may touch upon beam stability, head erosion dynamics, and wakefield excitation efficiency.  The evolution of partially bunched systems through the various mechanisms discussed here provide an ideal platform for investigating beam seeded self-modulation in plasmas. It also provides a sensitive system to investigate head erosion and methods for its mitigation, including an experimental comparison between self-ionized and laser-ionized performance \cite{head_solution}.

Beyond quasi-nonlinear resonance occurring when the beam is micro-bunched at a spacing of $\lambda_p$, more sophisticated schemes may be tested. One such scenario utilizes a linearly-ramped beam current that is micro-bunched at spacing of 1.5$\lambda_p$. This scheme may permit a large transformer ratio to be reached \cite{linearramp}. It would, however, enhance ion collapse and  could lead to instabilities due to self modulation of the beam structure at $\lambda_p$. Such scenarios will be investigated theoretically to evaluate the feasibility of their implementation. 

The studies presented here have assumed bunching with a near-IR (2 $\mu$m wavelength) laser. This choice was motivated by the desire to access TV/m-class fields using a resonantly excited PWFA system.  This choice in turn places stringent demands on beam quality, and scales all parameters involved in the experiment downward in size -- notably the transverse emittance and beam sizes. Further, operation with gas and attendant plasma densities close to cutoff for ionizing lasers would introduce challenging focal and propagation effects should a laser be used in the experiment. To mitigate experimental challenges introduced by this scaling, one may use a longer wavelength laser, {\sl e.g.} 10 $\mu$m, permitting higher emittance, higher charge beams to be used, while employing notably lower plasma density. This choice would lower the field expected to the few 100 GV/m range, which are still of high interest. 

In conclusion, the resonant excitation of PWFA with an optical-IR period micro-bunch train promises to be a robust alternative for accessing TV/m plasma wakefields. This initiative takes advantage of recent experimental progress in micro-bunch creation in high brightness electron beams at multi-GeV energy. Further, with small modifications to existing infrastructure at FACET-II, the experimental scenarios relevant to E-317 described here will be enabled. In this regard, we note that the use of micro-bunched beams with attosecond structure are also needed for E-318, an ultra-fast atomic physics experiment planned for FACET-II. This experiment, which can uniquely explore atomic electron dynamics under the influence of TV/m unipolar electric fields, is highly synergistic with the PWFA experiments described above, exploring effects which may be exploited as diagnostics in E-317.   

\section*{ACKNOWLEDGMENTS}

This work was performed with support of the US Dept. of Energy, Division of High Energy Physics, under contract no. DE-SC0009914.

\bibliography{apssamp}
\end{document}